\documentclass[aps,prb, twocolumn,nofootinbib]{revtex4}
\usepackage{longtable}
\usepackage{subfigure}
\usepackage{amsmath,amssymb,amsfonts,bm}
\usepackage[dvips]{color}
\usepackage{lscape, graphicx}
\def\KeyWord#1{$\backslash$\IfColor{$\!\!$\textRed{#1}\textBlack}{#1}$\!\!$}

\newcommand{\w}{ \omega }
\newcommand{\p}{\psi }

\newcommand{\be}{\begin{equation} }
\newcommand{\ee}{\end{equation} }
\newcommand{\ba}{\begin{eqnarray} }
\newcommand{\ea}{\end{eqnarray} }
\newcommand{\n}{\nonumber \\ }
\newcommand{\co}{\chi_{ij}^{(0)}}

\newcommand{\bk}{\mathbf{k}}

\newcommand{\bR}{\mathbf{R}}

\begin{document}
\input{epsf}
\title{Monopole Flux State on the Pyrochlore Lattice }
\author{F. J. Burnell}
\address{Department of Physics, Princeton University, Princeton, NJ 08544}
\author{ Shoibal Chakravarty}
\address{ Princeton Environmental Institute, Princeton University, Princeton, NJ 08544}
\address{Department of Physics, Princeton University, Princeton, NJ 08544}
\author{ S. L. Sondhi}
\address{Princeton Center for Theoretical Science, Princeton University, Princeton, NJ 08544}
\address{Department of Physics, Princeton University, Princeton, NJ 08544}
\date{\today}

\begin{abstract}
The ground state of a spin $\frac{1}{2}$ nearest neighbor quantum
Heisenberg antiferromagnet on the pyrochlore lattice is
investigated using a large $N$ $SU(N)$ fermionic mean field
theory. We find several mean field states, of which the
state of lowest energy upon Gutzwiller projection, is a parity and time
reversal breaking chiral phase with a unit monopole flux exiting each
tetrahedron. This ``monopole flux'' state has a Fermi surface
consisting of 4 lines intersecting at a point. At mean field the
low-energy excitations about the Fermi surface are gapless spinons.
An analysis using the projective symmetry group of this state suggests
that the state is stable to small fluctuations which neither induce a
gap, nor alter the unusual Fermi surface.
\end{abstract}

\maketitle

\section{Introduction}

This paper lies at the intersection of two streams of research in contemporary quantum
magnetism---the study of spin liquids and the study of geometrically frustrated magnetism.
Specifically, we are interested in $S=1/2$ Heisenberg models on the pyrochlore lattice
and were motivated by asking whether they support a zero temperature phase that  breaks
no symmetries of the problem--- a fully symmetric quantum spin liquid.

The study of quantum spin liquids---being defined broadly as states of spin systems that do not exhibit long range Ne\'{e}l order down to zero temperature---iss currently in the midst
of a significant revival. The subject itself is decades old with its contemporary study
tracing its origins to Anderson's introduction of the resonating valence bond (RVB)
state \cite{pwa-rvb} and then to his suggestion\cite{pwa-hightc}, upon the discovery of the cuprate superconductors, that their behavior
was traceable to a parent spin liquid state. But its current vogue has much to do with
recent progress in constructing actual models that realize spin liquid behavior\footnote{For an introduction to this area see the recent Les Houches lectures by Misguich \cite{misguichleshouches} and the older review article Ref.~\onlinecite{misguichlhuillier}.}
and the recognition that a large class of spin liquids exemplify ordering beyond the broken
symmetry paradigm---they give rise to low energy gauge fields but not order parameters. That
such ``topological phases'' \footnote{Here we use the term ``topological phases'' in the looser sense of any phase with emergent gauge fields. Strictly speaking the term should be reserved for cases where the low energy gauge theory is a purely topological gauge theory.} also underlie a fascinating approach to
quantum computation \cite{topcompreview} only multiplies their interest.

The study of geometrically frustrated magnets \cite{rm-review} has intertwined roots.
Indeed, Anderson's 1972 paper identified a small value of the spin and geometric frustration
as two sources of quantum fluctuations that could favor a spin liquid. In recent years there
has been steady progress in both understanding the behavior of many geometrically frustrated
magnets but, more importantly, in synthesizing an increasing number of compounds that realize
challenging idealizations to increasing accuracy \cite{ramirez,kagome} leading to a resurgence
of interest in these systems as well.

The pyrochlore lattice is a natural object of study in this context. It is highly frustrated
and frequently realized as a sublattice of the spinels or the pyrochlores.
Potentially, it could host a spin liquid in $d=3$ for small values of the spin.
Much work has
gone into studying its magnetic properties in various contexts. Most notably, it is known
to lack long range order with nearest neighbor interacting classical spins \cite{chalk-moess}
but instead to exhibit an emergent gauge field and dipolar correlations as $T \rightarrow 0$.
[Interestingly, this physics is realized in the Ising ``spin ice" systems (Dy and Ho titanate)
\cite{spinice} although with an additional fundamental dipolar interaction that leads to further elegant
physics involving magnetic monopoles\cite{cms2008}.] Attempts to work about the classical limit, in the spin wave ($1/S$) expansion
have lead to some insight into the quantum ``order by disorder'' selection mechanism in this
limit. While the fate of the $1/S$ expansion is not settled \cite{henleyuzi}, there is however little reason to
think that it can be informative when it comes to small values of spin, especially the
$S=1/2$ case\footnote{This is as good a place as any to note that there is not,
to date, a good experimental $S=1/2$ antiferromagnet on the pyrochlore lattice.} that is our concern in this paper. This is so partly because the selection mechanism
at large $S$ is weak and leads to somewhat ornate states but also for the well-understood
reason that it misses out on tunneling processes that are sensitive to the Berry phases entering
the exact path integral \cite{henleyvondelft, olegspinliquid}.

Consequently, various authors have attempted to directly tackle the $S=1/2$ problem. Harris,
Berlinsky and Bruder \cite{HBB} initiated a cluster treatment in which the pyrochlore lattice is first
decoupled into, say, its up tetrahedra and then perturbatively reconnected. Subsequently
Tsunetsugu \cite{Tsunetsugu} worked out a more complete treatment along the same lines and found a dimerized state
with a four sublattice structure. The criticism that this work predicts symmetry breaking that
is put in at the first step has attracted a potential rebuttal in the work of Berg et al \cite{berg} with
the ``Contractor Renormalization'' or CORE technique. An alternative perspective on this physics was provided in \cite{msg} where
it was shown that an $SU(N)$ deformation produces a quantum dimer model whose physics is very
reminiscent of the HBB scenario. Unfortunately, the $N=2$ limit is manifestly problematic so it
has not been possible to declare victory in this work. Yet another attack on the problem \cite{tms} used
an alternative large N theory---equivalent to Schwinger boson mean field
theory ---and
found a delicate energetics at small values of spin (or boson density) which nevertheless strongly
indicated that the spin 1/2 problem must break {\it some} symmetry.\footnote{Ref.~\onlinecite{tms} shows that at asymptotically small boson densities the system {\it must} break some symmetry. The minor caveat is that this does not rule out a different solution intervening right near $S=1/2$.}

With this set of predictions of symmetry breaking as background, in the present work we bring another
approximate large N technique---that of ``slave fermions'' \cite{bza,AffleckMarston}---to bear on
the pyrochlore problem with a view to examining whether it produces a symmetric, spin liquid, alternative.
To this end we enumerate various translationally invariant mean field solutions of which the lowest
energy non-dimerized solution is one we call a ``monopole flux'' state; upon Gutzwiller projection
it also improves upon the fully dimerized states. While this state does not
break lattice symmetries in the manner of the HBB scenario, it is not a spin liquid in the sense
of breaking no symmetries at all. Instead it is a chiral spin liquid \cite{kalmeyerlaughlin,WWZ} and breaks parity (P) and
time reversal (T) symmetries. It also exhibits spinons in its mean field spectrum. We describe the
unusual mean field spectrum---which yields a Fermi surface consisting of four lines intersecting at a
point---and its low energy limit in some detail. This state was first reported in Ref.~\onlinecite{shoibalthesis}. Subsequently it  sparked a larger
investigation by R. Shankar and two of us \cite{Shankar} on flux Hamiltonians on root lattices of
Lie groups with minuscule decorations and these results were announced there previously. The stability
of the mean field structure to fluctuations is the next question of interest. We make progress in
that direction by enumerating the projective symmetry group (PSG) \cite{WenPSG} of the state and
showing that it forbids any terms that would destabilize the mean field Fermi surface. This still leaves
the fate of the gauge fluctuations open as a matter of dynamics and we expect to discuss this
elsewhere \cite{wip}. Finally we note that as we were finalizing this paper there appeared an independent evaluation of energies for Gutzwiller projected wavefunctions on the pyrochlore \cite{KimHan} which agrees with our results on that score.

And now to the organization of the paper. We begin with a brief overview of the large-N/mean field slave fermion treatment of the Heisenberg model in Section \ref{LargeN}.
In Section \ref{MFPyro} we apply this technique to generate several mean field
ans\"{a}tze on the pyrochlore lattice.  We identify the lowest energy
state, or monopole flux state, and discuss its interesting
properties.  Section \ref{PSGStab} reviews in general terms how the PSG protects a mean field state against developing symmetry-breaking terms.  The PSG derived arguments
for the stability of the monopole flux state are given in Section \ref{ThePSGSec}, where we derive the general
form of the symmetry permitted perturbations to the Hamiltonian.  We conclude in Section \ref{Conc}.  Details of the PSG for the monopole flux state can be found in Appendix \ref{PSG} while Appendix \ref{Gutz} explains the numerical technique used to carry out Gutzwiller projection.

\section{The Large-N Heisenberg Model: Spinons and Gauge Fields} \label{LargeN}

In this section we briefly review the large $N$ fermionic approach to the $S=1/2$
$SU(2)$ Heisenberg model which began as a mean field theory introduced by Baskaran,
Zou and Anderson \cite{bza} and was shortly thereafter systematized
via a generalization to $SU(N)$ by Affleck and Marston \cite{AffleckMarston}.

In this approach,  we first replace the bosonic spin operators of the
Heisenberg Hamiltonian
\begin{equation} \label{Heisenberg}
H= J \sum_{<ij>} {\bf S}_i \cdot \bf{S}_j
\end{equation}
with bilinears in fermionic ``spinon'' operators:
\begin{equation} \label{spins}
{\bf S}_i =\frac{1}{2} \sum_{\alpha, \beta} c^{\dag}_{i \alpha} {\bf \sigma}_{\alpha \beta} c_{i \beta} \ .
\end{equation}
The resulting Hamiltonian conserves the number of fermions at each site and the starting
spin Hamiltonian is recovered if we limit ourselves to physical states with exactly
1 particle per site. Up to a constant in the subspace of physical states, it can be re-written
in the suggestive form,
\begin{equation} \label{Heisenberg2}
H= - \frac{J}{2}  \sum_{<ij>}  \sum_{\alpha}  \sum_{\beta} c^\dag_{i \alpha} c_{j \alpha} c^\dag_{j \beta}
c_{i \beta}
\end{equation}
A mean field theory arises upon performing the Hubbard-Stratonovich decoupling
\begin{equation} \label{eq:decoupledH}
H= -\sum_{\alpha}  \sum_{<ij>} (c^\dag_{i \alpha} c_{j \alpha} \chi_{ij} + h.c.) + \frac{2}{J} \sum_{<ij>} |\chi_{ij}|^2
\end{equation}
and locally minimizing the classical field $\chi_{ij}$ to obtain self-consistency.

In order to understand
the nature of fluctuations about such mean field solutions it is conceptually convenient to consider the
path integral defined by the equivalent Lagrangian:
\begin{eqnarray} \label{Lag}
L& =& \sum_{i, \alpha} c^\dag_{i, \alpha} (i \partial_t+ \mu) c_{i, \alpha} + \sum_{i, \alpha} \phi_i (c^\dag _{i, \alpha} c_{i, \alpha} -1) \nonumber \\
 & &+\sum_{<ij>}\left [ \sum_\alpha (c^\dag_{i \alpha} c_{j \alpha} \chi_{ij} + h.c.) - \frac{2}{J} |\chi_{ij}|^2 \right ]
\end{eqnarray}
where $\phi$  is a Lagrange multiplier field enforcing
the single occupancy constraint $\sum_\alpha c^\dag_{i\alpha} c_{i\alpha}=1$.

The above Lagrangian (\ref{Lag}) is invariant under the local gauge transformations
\ba \label{GaugeTrans}
c^\dag_i \rightarrow c^\dag_i e^{-i \theta_i} \n
\chi_{ij} \rightarrow \chi_{ij} e^{i (\theta_i -\theta_j)} \n
\phi \rightarrow \phi + \partial \theta/ \partial t
\ea
which arise from the local constraints in the fermionic formulation. It follows that we have
reformulated the Heisenberg model as a problem of fermions that live on the sites of the original
lattice coupled to a $U(1)$ gauge field and an amplitude field (the phase and amplitude of $\chi_{ij}$)
that both live on the links of the lattice.  In other words, we may write $\chi_{ij} = \rho_{ij} e^{i a_{ij}}$, where $a_{ij} \rightarrow a_{ij} + \theta_i -\theta_j$ under the gauge transformation (\ref{GaugeTrans}). The mean field theory consists of searching for a saddle
point with frozen link fields.

As  the Lagrangian (\ref{Lag}) does not directly constrain the phase of the $\chi_{ij}$, it describes
a strongly coupled gauge
theory where the assumption of a weakly fluctuating gauge field invoked in the mean field theory
is, {\it prima facie}, suspect.
To circumvent this barrier, Affleck and Marston \cite{AffleckMarston} proposed a
large $N$ framework which introduces a weak-coupling limit for
the model (\ref{Lag}) by extending the SU(2) spin symmetry group
of the Heisenberg model to SU(N) with N even.  The result is a theory of
many spin flavors whose coupling strength scales as $J
\rightarrow J/N$.  In the limit that $N\rightarrow \infty$, the
corresponding mean field theory
 is exact; for sufficiently large but finite $N$ one hopes that a
perturbative expansion gives accurate results.
The validity of
the qualitative features deduced at large $N$ in the starting $SU(2)$
problem is, of course, hard to establish by such considerations
and requires direct numerical or experimental confirmation.

To effect the large N generalization, we
replace the 2 spinon operators $c_{\uparrow}$ and
$c_{\downarrow}$ with $N$ spinon operators $c_{\alpha}$. The
single occupancy constraint is now modified to
\begin{equation}\label{constr}
\sum_{\alpha=1}^N c^\dag_{i\alpha} c_{i \alpha} = \frac{N}{2}
\end{equation}
and the large-N Hamiltonian has the form
\begin{eqnarray} \label{NHam}
H&=&J/N \sum_{\alpha, \beta} \sum_{<ij>}c^\dag_{j \alpha} c_{i \alpha} c^{\dag}_{i \beta} c_{j \beta}\nonumber \\
&=&  -\sum_{\alpha}\sum_{<ij>} (c^\dag_{i \alpha} c_{j \alpha} \chi_{ij} + h.c.) + \frac{N}{J} \sum_{<ij>} |\chi_{ij}|^2
\end{eqnarray}
In the infinite N limit, the action is constrained to its saddle point
and the mean field solution becomes exact.
Further, to lowest order in $\frac{1}{N}$ the allowed
fluctuations involve moving single spinons, so that as $N\rightarrow \infty$
we need only impose the constraint (\ref{constr}) on average.

Away from $N=\infty$ the link fields, especially the gauge field, can fluctuate again although now
with a controllably small coupling. While the fate of the coupled fermion-gauge system still
needs investigation, the presence of a small parameter is a great aid in the analysis, as in the
recent work on algebraic spin liquids \cite{HermeleAlg}.

Finally, we note that the starting $SU(2)$ problem is special, in that it is naturally formulated as
an $SU(2)$ gauge theory \cite{afflecketal,dagottoetal}. This can have the consequence that the $N=2$
descendant of the large $N$ state, if stable, may exhibit a weakly fluctuating $SU(2)$ gauge field instead of the
$U(1)$ field that arises in the above description. We will comment on this in the context of this
paper at the end.

\section{Mean-Field Analysis} \label{MFPyro}

\subsection{Saddle Points of the Nearest Neighbor Heisenberg Model}

We begin by enumerating mean field (MF) states which preserve translation invariance
on the pyrochlore.
A mean field solution consists of a choice of link fields which minimizes the
mean field energy functional for the Lagrangian (\ref{Lag})
\be \label{EMF}
E(\chi) = N\left [ \sum_{ \langle ij \rangle} \frac{1}{J}
|\chi_{ij}|^2 + \sum_\bk
  (\varepsilon (\bk) - \mu) \right ]
\ee
where $\varepsilon (\bk)$ is the energy of a spinon of momentum $\bk$ in the
fixed background $\chi_{ij}$, and the chemical potential $\mu$ is chosen
so that
the constraint of 1 particle per site is satisfied on average.

As discussed in the introduction, previous work on the Heisenberg model on the pyrochlore lattice has led to ground states
with broken symmetries.  In this work we are particularly interested in constructing a natural state on the pyrochlore that
breaks as few symmetries as possible.  To this end, we begin our search with especially symmetric {\it ans\"{a}tze} for which
$\rho_{ij} \equiv \rho $ is independent of $i$ and $j$, and the flux $\Phi_\triangle = \sum_{\triangle} a_{ij}$ through each
face of the tetrahedron is the same.  The net flux $\sum_{i=1}^4 \Phi_\triangle$ through each tetrahedron must be an integer
multiple of $2 \pi$, since each edge borders two faces such that its net contribution to the flux is 0 (mod $2\pi$).
This gives the following $3$ candidate spin liquid states:
\begin{enumerate}
\item Uniform: $\Phi_{\triangle} =0$
\item $\pi$ Flux: $\Phi_{\triangle} =\pi$
\item Monopole: $\Phi_{\triangle} =\pi/2$.  Every triangular face of the
tetrahedron has a $\pi/2$ outwards flux -- equivalent to a monopole
of strength $2\pi$ placed at the center of each tetrahedron.
\end{enumerate}
At infinite $N$ a dimerized state is always the global minimum of (\ref{EMF})
\cite{Dimers}; thus we also consider
\begin{enumerate}
\addtocounter{enumi}{3}
\item Dimerized:  $\chi_{ij} = \chi^0$ on a set of bonds that constitute any dimer covering of the lattice but zero otherwise.
\end{enumerate}

The states (1-3) above are analogues of the uniform, $\pi$ flux, and chiral states
studied previously on the square lattice \cite{AffleckMarston, WWZ}.  Of the above states, $(1)$ and $(2)$ break no symmetries of
the problem; the third preserves lattice symmetries but breaks P and T.

The states $(1)$ and $(2)$ are in fact particle-hole conjugates: a particle-hole transformation maps $c^\dag_i c_j + c^\dag_j c_i \rightarrow -c^\dag_j c_i - c^\dag_i c_j$, changing the sign of $\chi$ on each bond and adding $\pi$ flux to each triangular plaquette.  At $N=2$ this can be effectuated by an $SU(2)$ gauge transformation, so that the states $(1)$ and $(2)$ describe the same state after Gutzwiller projection.

The mean field energies of these states are listed in the first column of Table I. Consistent with Rokhsar's general
considerations \cite{Rokhsar,Dimers} the fully dimerized state is lowest energy and the monopole flux state has the lowest
energy of the non-dimerized states. The mean field states with N set equal to 2 do not satisfy the
single occupancy constraint. While, in principle, perturbation theory in $1/N$ can greatly improve the wavefunction
in this regard this is a complex business (to which we return in Sections IV and V) ill-suited to actual energetics.
Instead, the somewhat {\it ad hoc} procedure of (Gutzwiller) projecting the mean field wave function
onto the Hilbert space of singly occupied sites is typically employed to improve matters. This leads to resonances and
long range correlations that can substantially lower the mean field ground state energy, particularly for spin-liquid type
states.

Expectation values in the Gutzwiller projection of a state can be carried out using a Monte Carlo
approach, as described in Ref.~\onlinecite{gros}.  A brief description of the numerical method specialized to our problem
is given in Appendix \ref{Gutz}. The second column in Table \ref{TabEnergies} shows the numerically evaluated
energies of the 4 mean field states with Gutzwiller projection. We see that the monopole flux state now emerges
as the lowest energy state of our quartet. Encouraged by this, and also because the state has various elegant
properties, we will focus in the remainder of this work on the properties of the monopole flux state. Note however,
that we have failed to preserve all symmetries of the Hamiltonian even in this approach---we are forced to break
$T$ and $P$ and thus end up with a chiral spin liquid. We give a fuller description of the symmetries of the state
below.

Finally, we note that larger unit cells can be consistent with translationally invariant
states.\footnote{We are grateful to Michael Hermele for emphasizing this point.} Such states have an integral multiple of $\pi/2$ flux through
each triangular plaquette, but also non-trivial flux through the hexagonal plaquettes in
the kagom\'{e} planes, as for the mean field states on the kagom\'{e} studied in
Ref.~\onlinecite{HermeleKagome}.  By the same arguments as employed for a single tetrahedron,
we find that the flux through the hexagons must have values $0$ or $\pi$ (mod $2\pi$) to
preserve the translational symmetry of the lattice.  [A flux of $\pi/2$ per hexagonal plaquette necessarily breaks lattice translations].  However, as noted in Table \ref{TabEnergies}, we find that these states also have higher energies than the monopole flux state both at mean field and upon Gutzwiller projection.
\begin{table}
\begin{center}
\begin{tabular}{|c|c|c|}
\hline
&  $E_{MF}$ (unprojected) & $E_{MF}$ (projected) \\
\hline
Uniform &$-0.3333 J$  & $-.3752 \pm 0.0004$ \\
$\pi$ Flux & $-0.3333J$  &  $-.3752 \pm 0.0004$  \\
Monopole & $-0.3550 J $&  $-.4473 \pm 0.0009$  \\
Dimer & $-.375 J$ & $-.375 J$ \\
($\pi, \pi)$ &$-0.3333 J$ & $-0.3751 \pm 0.0008 $\\
$(\pi/2, \pi)$ &  $-0.3491J $ & $-0.4353 \pm 0.002$\\
\hline
\end{tabular}
\caption{Mean-field energies for projected and unprojected ground
states of the mean field {\it ans\"{a}tze} considered.  The quoted mean field energies are the energy of (\ref{eq:decoupledH}) plus the omitted constant $-\frac{J}{4}$ per site required to make a correspondence with (\ref{Heisenberg}).  The states $(\pi, \pi)$ and $(\pi/2, \pi)$ are variants of the uniform and monopole state, respectively, with flux $\pi$ per hexagonal plaquette.  The projected wave functions were evaluated on a lattice of $5 \times 5 \times 5$ unit cells, or 500 sites for configurations with a 4 site unit cell, and 1000 sites for configurations with an 8 site unit cell.}
\label{TabEnergies}
\end{center}
\end{table}

\subsection{\label{Monpole}The Monopole Flux State}

\begin{figure}[htb]
\begin{center}
\epsfxsize=2.5in
\epsffile{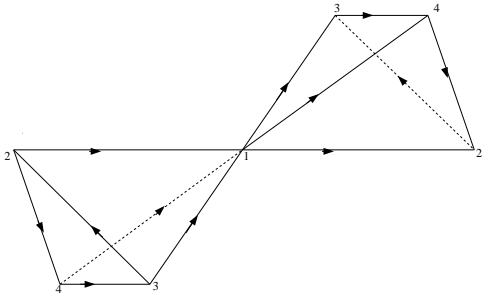}
\caption[Bond Order in The Monopole Phase]{Link field orientations in the  monopole
  state  with $\chi=\pm i \rho_0 $.  Hopping along the direction of an arrow induces a phase of $\pi/2$; hopping against the arrows, a phase of $- \pi/2$.  The flux on each triangular face is $\pi/2$ outwards.  With this flux assignment the monopole flux state breaks $T$ and $P$, but is invariant under lattice translations and rotations.} \label{monopolefig}
\end{center}
\end{figure}

The monopole flux state exhibits a flux
of $\pi/2$ per triangular face.  To write down the mean field Hamiltonian explicitly
we must pick a gauge.  We choose
$\chi_{ij} = \rho_0 e^{i a_{ij}}$, with $a_{ij} = \pm \pi/2$.
The phase of $\pm i$ that a spinon picks up when hopping from site
$i$ to $j$ can be represented as an arrow on the corresponding edge,
which points from $i$ to $j$ ($j$ to $i$) if the resulting phase is
$+(-) i$.
The orientation of the link
fields, shown in Figure \ref{monopolefig}, gives an outward flux of $\pi/2$
per plaquette.

The necessity of picking a gauge for the mean field solution causes, as usual,
various symmetries to be implemented projectively.
For example, the assignment shown in Figure \ref{monopolefig} is not invariant under lattice rotations.  However, the background link fields after
rotation can be gauge transformed to the original state, as expected from the manifestly rotation invariant assignment of fluxes.  We discuss
these and other symmetries in more detail in
Sections \ref{PSGStab} and \ref{ThePSGSec}; here we merely note that $P$ and $T$ are the only symmetries broken by the monopole flux state.

The Hamiltonian for spinons in the gauge choice shown in Fig. \ref{monopolefig} is
\begin{widetext}
\begin{eqnarray}\label{hmonopole}
H&
=&-\frac{2N\rho}{J}\sum_{\bm{k},\alpha}\Psi^{\dagger}_{\bm{k} \alpha}
\begin{bmatrix}
0                &     \sin(\frac{k_x+k_y}{4})          &     \sin(\frac{k_y+k_z}{4})     &      \sin(\frac{k_x+k_z}{4})\\
\sin(\frac{k_x+k_y}{4})   &  0                   &   \sin(\frac{k_x-k_z}{4})         &  \sin(\frac{k_z-k_y}{4})\\
\sin(\frac{k_y+k_z}{4})& \sin(\frac{k_x-k_z}{4})&         0              &      \sin(\frac{k_y-k_x}{4})\\
\sin(\frac{k_x+k_z}{4})& \sin (\frac{k_z-k_y}{4}) &  \sin (\frac{k_y-k_x}{4}) &         0
\end{bmatrix}
\Psi_{\bm{k}\alpha}.
\end{eqnarray}
\end{widetext}
where $\Psi$ is a 4-component vector, with $\Psi_{\bm{k} \alpha}^{i} = c^{ i}_{\alpha \bm{k}}$.  Here the index $i$ labels
the 4 sites in the tetrahedral unit cell.

\begin{figure}[ht]
\centering
\subfigure[] 
{
    \includegraphics[width=2.75in]{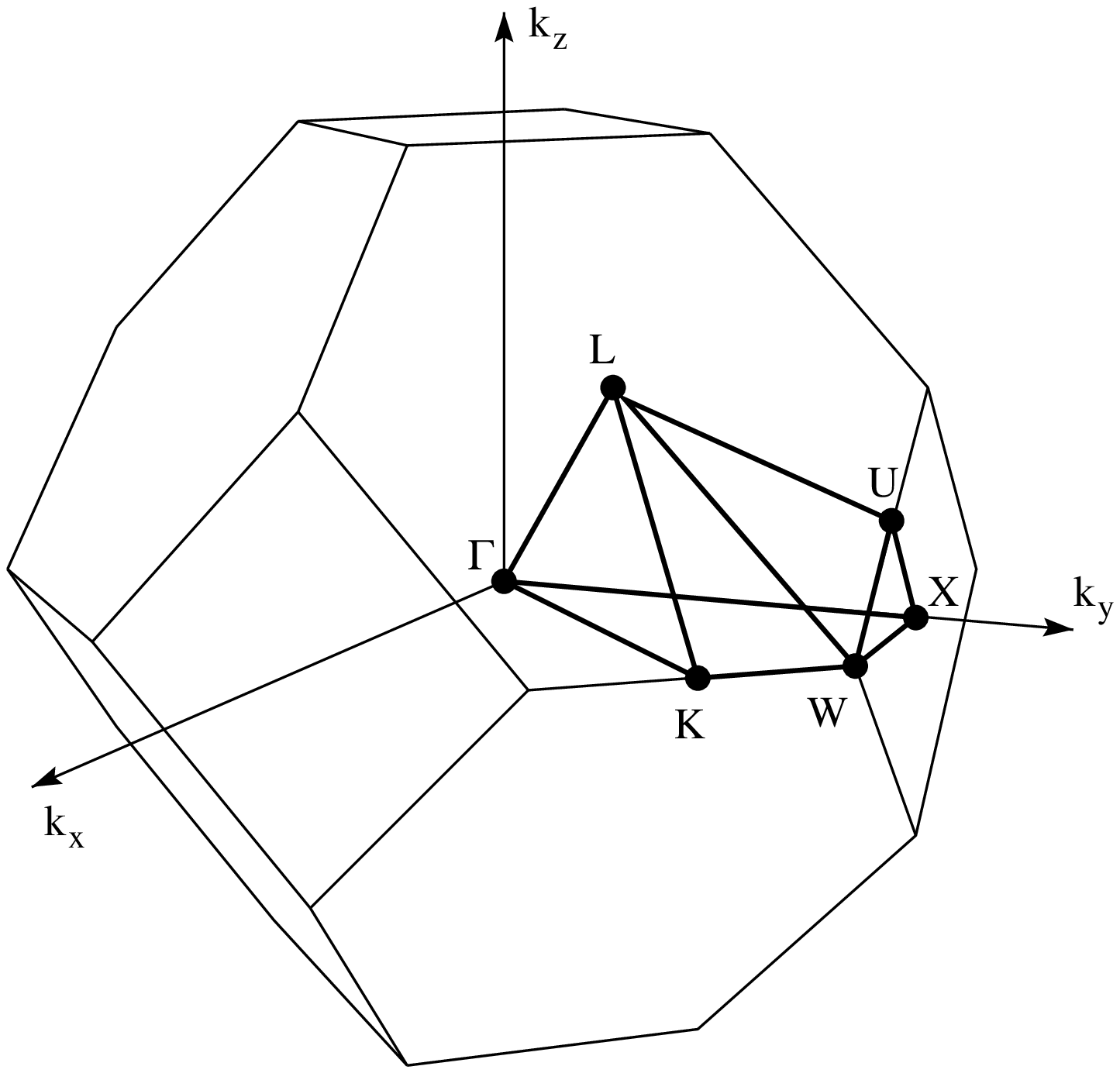}
    \label{Zone}
} \hspace{1cm}
\subfigure[] 
{
    \includegraphics[width=2.75in]{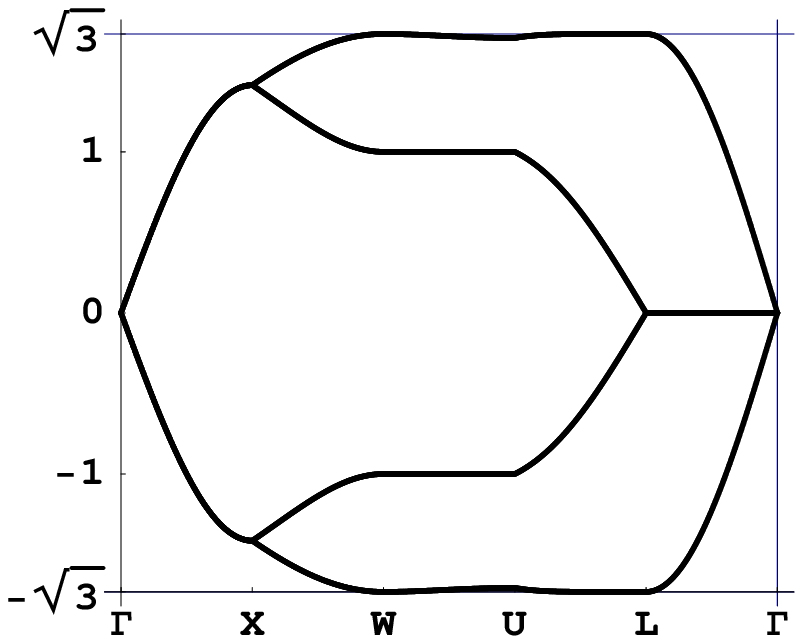}
    \label{mono}
} \hspace{1cm}
\caption[]{(b) shows the spectrum of the monopole flux state. Note
the Fermi line $\frac{1}{\sqrt{3}}(k,k,k)$.  (a) shows the contour in the Brillouin
zone along which the spectrum is plotted. }
 \end{figure}
Figure \ref{mono} shows a plot of the energy eigenvalues of
(\ref{hmonopole}) along the high-symmetry lines of the Brillouin
zone.  At half filling, the Fermi `surface' consists of the lines
$k(\pm 1,\pm 1,\pm 1)$ which join the point $(0,0,0)$ to the
center of the hexagonal faces of the  Brillouin zone of the cubic
FCC lattice, line $(L-\Gamma)$ in  Fig. \ref{mono}. Each Fermi
line has a pair of zero energy eigenstates.

\begin{figure}[ht]
\centering
\subfigure[] 
{
    \includegraphics[width=5.9cm]{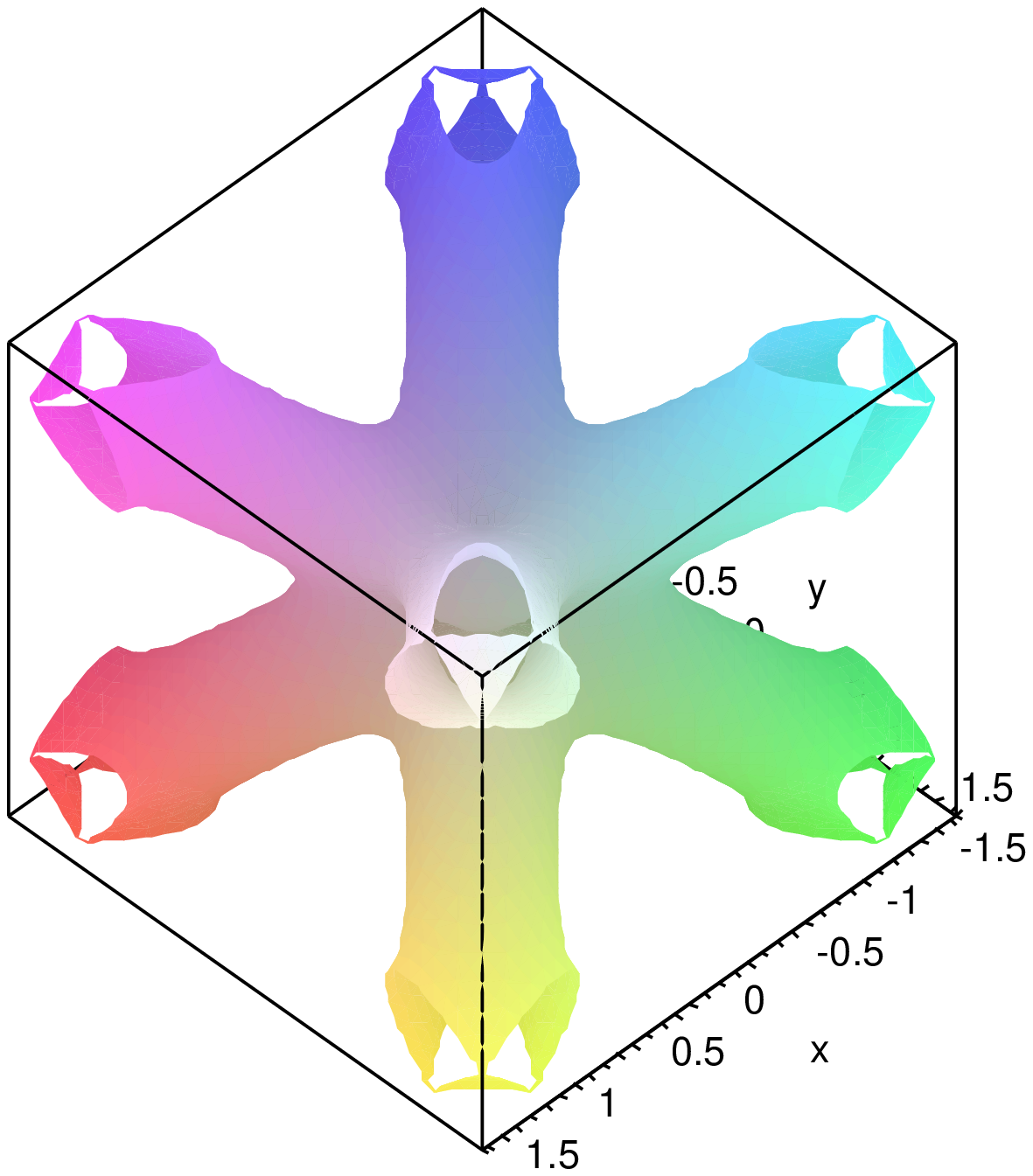}
    \label{fsa}
} \hspace{1cm}
\subfigure[] 
{
    \includegraphics[width=5.9cm]{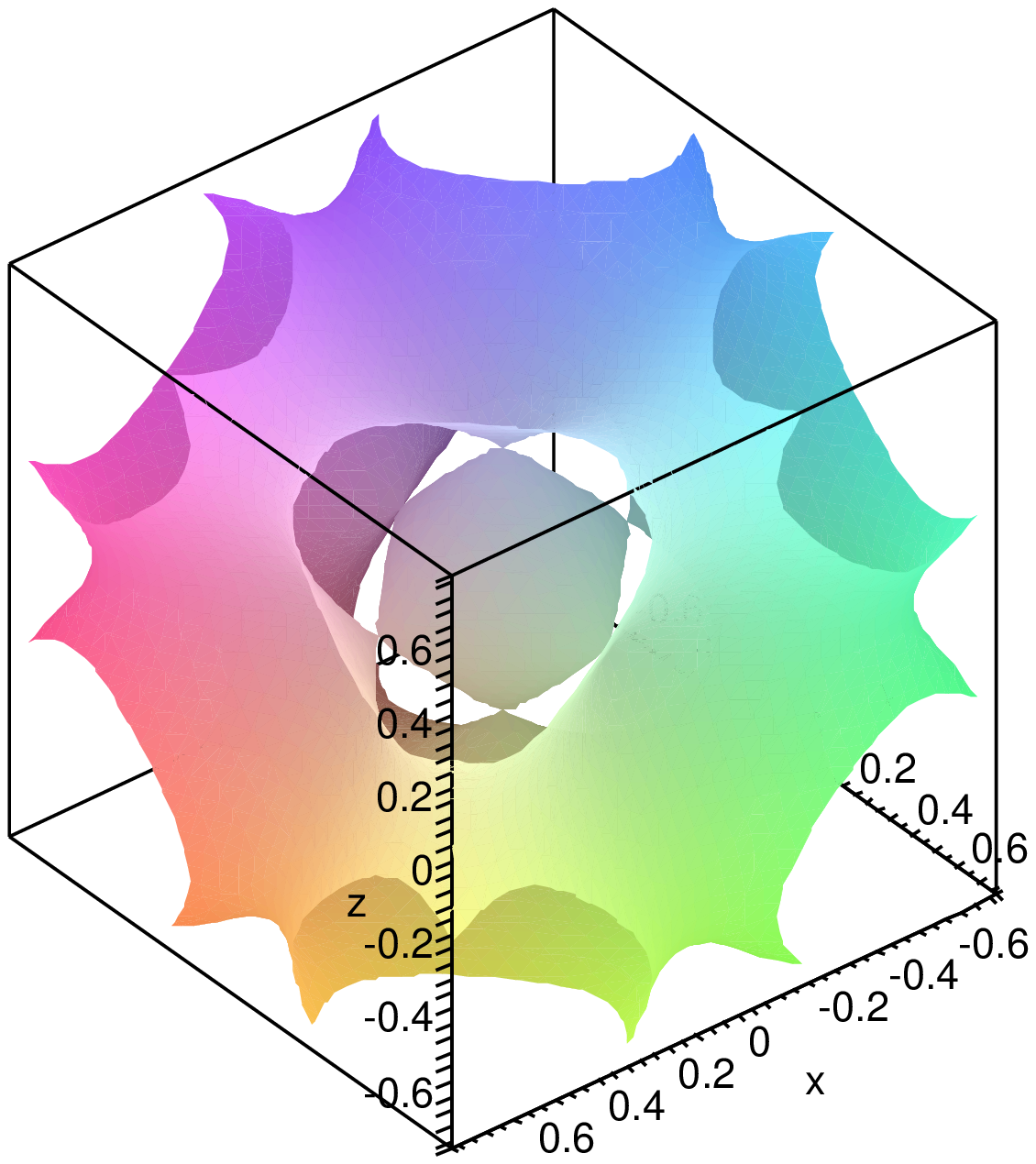}
        \label{fsb}
 }
\caption[]{Constant energy surfaces of the monopole flux state, for
$E/J =0.5$.  Decreasing $E/J$ makes the cylinders thinner.
(a) A view of the cube of side length $\pi$ surrounding the
origin.  Eight cylinders surrounding the eight Fermi lines
emanate from the origin; at the origin a diamond-line shape (the
low-energy spectrum of the remaining two bands) can also be
seen.  This shape repeats at the cube´s corners ($\pm \pi, \pm
\pi$). (b)  A close-up view of the region surrounding the
origin.  Altering the ratio $E/J$ shrinks the entire structure,
but does not change its shape.  }
\end{figure}

Figures \ref{fsa} shows a surface of constant energy $E \approx
0$ near the Fermi surface.  At $E=0$, the 4 bands intersect only
at the origin and the constant energy surface is given by the 4
lines described above.  Surfaces of constant energy $E \approx 0,
E \neq 0$ consist of 4 cylinders enclosing the $(1,1,1)$ directions, which
are the surfaces of constant energy for particle-like ($E>0)$ or hole-like
$(E<0)$ excitations
about the Fermi line.  About the origin all $4$ bands have energy
linear in $k$, and another, diamond-shaped constant-energy surface
appears.  These surfaces
intersect at the band crossings along the $x$, $y$, and $z$
axes\footnote{ This Fermi surface does not display fermion doubling
in the naive sense; all four bands cross at only one point in the
Brillouin zone.  This does not violate the result of Ref.~\onlinecite{Nielsen}, which
assumes that levels are degenerate only at a finite set of points.}.

\subsection{Low Energy Expansions of the Spinon Dispersion}

The low-energy structure of the monopole flux state can be divided into two regions:
$R_1$, the set of 4 Fermi lines sufficiently far from the origin, and
$R_0$,  the area near the origin.

In $R_1$, only 2 of the four bands lie near the Fermi surface, and the
low-energy theory is effectively two-dimensional.  Linearizing the
Hamiltonian about one of the Fermi lines gives:
\begin{equation} \label{Hline}
H[\varepsilon,\theta]=\Psi^{\dagger}_{1\alpha}[k,\varepsilon,\theta](\varepsilon\cos\theta\tau_1+\varepsilon\sin\theta\tau_2)\Psi_{1\alpha}[k,\varepsilon,\theta],
\end{equation}
with energies $\pm \varepsilon$,
independent of $\theta$.
Here we have used the local coordinate system
\begin{equation}
(q_1,q_2,q_3)=(k+\sqrt{\frac{2}{3}}u,k-\frac{u}{\sqrt{6}}-\frac{v}{\sqrt{2}},k-\frac{u}{\sqrt{6}}+\frac{v}{\sqrt{2}}),
\end{equation}
 with  $\theta(u,v)=\tan^{-1}(v/u)$ and $\epsilon(u,v)=\sqrt{u^2+v^2}/(2\sqrt{2})$.
Curiously, at mean field the low energy spectrum is independent of the
position $k$ {\it along} the Fermi line, depending only on the
momentum component in the kagom\'{e} planes perpendicular to the
vector $\bm{l_i}$.  Thus the linearized theory away from the
origin consists of a continuum of flavors of Dirac fermions confined to
the kagom{\'e} planes orthogonal to this line.

In $R_0$ all 4 bands have energies vanishing linearly as
$k\rightarrow 0$, and the low-energy Hamiltonian is given by:
\begin{equation} \label{Hlin}
H=-\frac{2N\rho}{J}\sum_{k,\alpha}\Psi^{\dagger\alpha}_{k}
\begin{bmatrix}
0                &     \frac{k_x+k_y}{4}          &     \frac{k_y+k_z}{4}    &      \frac{k_x+k_z}{4}\\
\frac{k_x+k_y}{4}   &  0                   &   \frac{k_x-k_z}{4}         & \frac{k_z-k_y}{4}\\
\frac{k_y+k_z}{4}& \frac{k_x-k_z}{4}&         0              &      \frac{k_y-k_x}{4}\\
\frac{k_x+k_z}{4}& \frac{k_z-k_y}{4} &  \frac{k_y-k_x}{4} &         0
\end{bmatrix}
\Psi_{k\alpha}.
\end{equation}
with energy eigenvalues
\begin{equation}
\varepsilon=\pm\sqrt{\frac{1}{8}\sum_i k_i^2\pm\frac{1}{8}\sqrt{3\sum_{(i< j)}k_i^2k_j^2}},
\end{equation}
This dispersion relation also gives massless spinons;
however, the theory is no longer one of Dirac fermions.

In addition to four bands touching at the origin, the linearized
Hamiltonian (\ref{Hlin}) has 2 zero eigenvalues on each Fermi line.
Restricting the spinors to the corresponding low-energy subspace again
yields the expansion (\ref{Hline}).
Thus (\ref{Hlin}) captures the principal features of the low-energy behavior not only in the vicinity of
the origin, but throughout the entire Brillouin zone.

The linearized Hamiltonian has several interesting features.
First, we may express it in terms of three matrices as follows:
\be H = \alpha_x k_x + \alpha_y k_y + \alpha_z k_z \ee The
$\alpha$ matrices are reminiscent of Dirac $\gamma$ matrices,
albeit with a tetrahedrally invariant, rather than rotationally
invariant, algebraic structure.  They do not comprise a Clifford
algebra, but obey the anti-commutation relations
 \ba \label{acom1}
\left\{ \alpha_i , \alpha_j \right \}&=& 2\delta_{ij} +
\sqrt{3}|\varepsilon_{ijk}|W_k  \n
 \left \{ W_i , W_j \right \}&=& 2\delta_{ij}
 \ea

Further, in a 3+1 dimensional Dirac theory there are two matrices
($\gamma_0$ and $\gamma_5$) which anti-commute with all $\gamma_i$.
In this sense our mean field Hamiltonian more resembles a 2+1 dimensional
Dirac theory: there is a unique matrix $\alpha_0$ such that $\{
\alpha_0, \alpha_i \}=0, i=1..3$, given by
\begin{widetext}
\be \label{alpha0}
\alpha^{(0)} =
\frac{1}{\sqrt{N}}\begin{bmatrix}
0 & \cos(y) \cos(x) & \cos(y)\cos(z)& \cos(x)\cos(z) \\
-\cos(y) \cos(x)& 0& \cos(x)\cos(z)& -\cos(y)\cos(z) \\
-\cos(y)\cos(z)& -\cos(x)\cos(z)& 0& \cos(y)\cos(x) \\
-\cos(x)\cos(z)& \cos(y)\cos(z)& -\cos(y)\cos(x)& 0 \\
\end{bmatrix}
\ee
\end{widetext}
where $N$ is a normalization factor such that $(\alpha^{(0)}) ^2= 1$.  In the continuum limit this reduces to
\be
\alpha^{(0)} =\begin{bmatrix}
0& \frac{1}{\sqrt{3}}& \frac{1}{\sqrt{3}}& \frac{1}{\sqrt{3}}\\
\frac{-1}{\sqrt{3}}&0&\frac{1}{\sqrt{3}}&\frac{-1}{\sqrt{3}} \\
\frac{-1}{\sqrt{3}} & \frac{-1}{\sqrt{3}}&0&\frac{1}{\sqrt{3}} \\
\frac{-1}{\sqrt{3}}& \frac{1}{\sqrt{3}}& \frac{-1}{\sqrt{3}}& 0 \\
\end{bmatrix}
\ee
$\alpha^{(0)}$ acts as a spectrum inverting operator on $H$,
interchanging hole states at energy $-E(k)$ with particle states at energy $E(k)$.

We point out that many of the interesting features of the low-energy spectrum of the monopole flux state
can be generalized to a class of lattices whose geometry is related to certain
representations of Lie groups \cite{Shankar}.  Indeed, the four sites in the
tetrahedral unit cell can be viewed as the four weights in the fundamental
representation of SU(4); hopping on the pyrochlore is then analogous to acting with
the appropriate raising and lowering operators.  This perspective gives an explicit
connection between the hopping Hamiltonian (\ref{Hlin}) and the ladder operators
in the fundamental representation of SU(4).  Analogous hopping problems can be
studied for various other Lie group representations, as outlined in detail in Ref.~\onlinecite{Shankar}.

To summarize, the monopole flux state is a spin liquid which preserves all
symmetries of the full Hamiltonian except $P$ and $T$.  At mean field
level it has gapless spinons along a 1-dimensional Fermi surface of $4$
lines which intersect at the origin.  Though strictly at $N = \infty$ it
has higher energy than the dimerized state, Gutzwiller projection suggests
that for $N=2$ this is no longer the case, and the monopole flux state is the lowest
energy simple mean field {\it ansatz} after projection.
 We now turn our
attention to what can be said about the stability of this rather
unusual mean field state.

\section{Stability of the Mean-Field solution: the role of the PSG} \label{PSGStab}

Next, we would like to address the question of whether the
mean field solutions described above maintain their basic
properties at finite $N$ and whether this holds all the way to $N=2$.
This is a difficult problem, whose complete solution is not available
even for the longer studied cases of the algebraic spin liquids in
$d=2$ \footnote{See Refs.~\onlinecite{HermeleAlg,HermeleMother} and references therein.} However, following that work the general idea
would be to try and understand if the state is truly stable at large
enough $N$ while leaving the question of stability at small $N$ to
detailed numerical investigation.

There are several questions here. First, is the mean field solution locally
stable? Second, is it the global minimum? Third, assuming the answer thus
far is in the affirmative, is the expansion about the mean field solution
well behaved? Ideally, this would mean convergent, but it would be sufficient
to know that it does not destroy the qualitative features of the gapless
spinon dispersion at mean field. For example, in the case of the algebraic
spin liquids in $d=2$ the spinons interact and acquire anomalous dimensions
away from $N = \infty$ but they remain gapless in the vicinity of a discrete
set of points \cite{HermeleAlg}.
Finally, what is the spectrum of collective (gauge) excitations
that arise in this expansion?

Based on the experience with spin liquids in $d=2$, answering the first
two questions in the affirmative is likely to require the addition of
more terms to the Hamiltonian although it may be possible to choose them so
that they become trivial at $N=2$ \cite{AffleckMarston}. We have not
investigated this in detail but there does not appear to be an obstacle to
doing this.

The third and fourth questions require detailed consideration of the symmetry
properties and the detailed dynamics of the expansion which is that of a lattice
gauge theory with matter and gauge fields in some fashion. In this work we will
carry out the first part of this program which goes under the study of the
``Projective Symmetry Group" (PSG) discussed in detail by Wen \cite{WenPSG}.
In this section we review the concept of the PSG and its implications for
perturbative expansions.  We also show that at $N= \infty$, or in mean field theory,
the PSG already helps us understand the stability of particular mean field
solutions; to our knowledge this particular aspect has not appeared in the
literature before.

Turning first to the PSG, observe that though
the original Hamiltonian formulated in terms of spin operators is
invariant under the full space group of the pyrochlore lattice, the actual
mean field Hamiltonian of the monopole flux state is not: many of the symmetry transformations
map the mean field Hamiltonian into different {\it but gauge equivalent} Hamiltonians.
Thus, when working in the gauge theory formulation of the problem,
the actual symmetry transformations  of the mean field  Hamiltonian have the form:
\be \label{PSGTrans1}
c_i \rightarrow g_s(s(c_i))
\ee
where $s$ is an element of the space group, and $g$ is a gauge
transformation.  As the full Hamiltonian is gauge invariant,
(\ref{PSGTrans1}) is simply an alternative formulation of the
lattice symmetries.  Hence as emphasized by Ref.~\onlinecite{WenPSG}, these projective
symmetry operators are exactly analogous to lattice symmetries in the
original spin problem.  Indeed, the correct choice of gauge
transformation ensures that both $H_{MF}$ and $H$ are invariant under
the PSG, so that the family
\be
H_\lambda = H_{MF} + \lambda (H - H_{MF} )
\ee
is also invariant and perturbative corrections in $H - H_{MF}$ cannot
break the PSG symmetry.

Before discussing the implications of PSG symmetry for the monopole flux state, we would like to briefly underline how the PSG
constrains the mean field theory at infinite $N$ which is a much simpler but
still instructive exercise.

Ignoring the dimerization instability, the monopole flux state is a mean field
minimum for nearest-neighbor couplings. The PSG is the symmetry group of
the corresponding mean field Hamiltonian. We may now ask what happens to the PSG
if further neighbor couplings are included in the Hamiltonian, in particular
do they lead to terms in the new mean field Hamiltonians that modify the PSG found
earlier?

At $N=\infty$ this is a problem of minimizing the expectation value of the sum
of the quadratic Hamiltonian in Eq.(\ref{eq:decoupledH}) and the new generic
terms
\begin{equation}
\delta H= - \sum_{\alpha}  \sum^{'}_{(ij)}
(c^\dag_{i \alpha} c_{j \alpha} \chi_{ij} + h.c.) + \frac{N}{J_{ij}}
\sum_{<ij>} |\chi_{ij}|^2
\end{equation}
wherein the primed sum runs over non-nearest neighbor bonds and the $J_{ij}$
are much smaller than the nearest neighbor $J$.
We will now show that, generically, the result of the new minimization  for
the perturbed problem preserves the PSG for the nearest neighbor problem.
While we use the language of perturbing about the monopole flux state, the
argument is general.

With the addition of the perturbation, the functional that we need to minimize
over the full set of $\{\chi_{ij} \}$ is:
\ba \label{deltaMF}
E_{MF}= \langle H + \delta H \rangle_{H + \delta H}  + \frac{N}{J}
\sum_{<ij>} |\chi_{ij}|^2 + \frac{N}{J_{ij}} \sum_{(ij)} |\chi_{ij}|^2 \ .
\ea
Let $\co$ denote the values of the link fields when $\delta H \equiv 0$,
i.e. in the monopole flux state.
For small $J_{ij}$ we expect the new minimum to lie not far from the old
one, whence the link fields will be close to the values $\co$. Consequently
we will compute the expectation value required in the above equation in
perturbation theory in $\delta H$ about $H$. (If such an expansion fails to have
any radius of convergence then we are already parked at a phase transition and
no stability argument is possible.)

This expansion,
\ba
E_{MF} (\{\chi_{ij} \}) = E_0  + \langle 0 | \delta H | 0 \rangle
+ \sum_{n >0} \frac{|\langle 0 | \delta H | n \rangle |^2}{E_0 - E_n}
+ \cdots \nonumber \\
+ \frac{N}{J}
\sum_{<ij>} |\chi_{ij}|^2 + \frac{N}{J_{ij}} \sum_{(ij)} |\chi_{ij}|^2 ,
\ea
where the numerical indices refer to the ground and excited states of the
unperturbed Hamiltonian $H(\{\co \})$, has three properties that we
need. First, the linear term takes the explicit form,
\ba
- \sum_{\alpha}  \sum_{(ij)} \langle c^\dag_{i \alpha} c_{j \alpha} \rangle \delta \chi_{ij}
+ h.c.
\ea
where $ \delta \chi_{ij}$ is $\chi_{ij}$ for the new bonds and the deviation
from $\co$ for the nearest neighbor bonds. This implies that new minimization
likes to turn on exactly those $\chi_{ij}$ that transform as the expectation
values $\langle c^\dag_{i \alpha} c_{j \alpha} \rangle $.
If these are, in fact, what get turned on, then the new mean field Hamiltonian
will indeed inherit the PSG of the starting one. The second property that
we need can be established by considering a decomposition of $\delta H$ into
a piece that commutes with the PSG generators and another piece that does not.
It is straightforward to see that terms from quadratic order and beyond must
give rise to a potential which is even in powers of the non-PSG conserving
piece of $\delta H$. Finally, at sufficiently small $J_{ij}$ the potential
for the $\chi_{ij}$ must be stable due to the explicit factors of $1/J_{ij}$.
Together these properties imply that the new minimum must be in the ``direction''
selected by the linear term and hence will exhibit the same PSG as before.

\section{The PSG of the monopole flux state} \label{ThePSGSec}

We will now describe the PSG of the monopole flux state, and its
implications for stability at the mean field level.
The space group of the pyrochlore lattice is $Fd\bar{3}m$, which contains
$24$ symmorphic and $24$ non-symmorphic elements.  For our purposes it is most convenient to divide these elements into the 24 proper elements composed of rotations and translations, and 24 improper elements involving a reflection or inversion.
The $24$ proper elements are:
\be
P^{0} = \{ 1, 8 C_3, 3 C_2, 6 \tilde{C}_2, 6 \tilde{C}_4 \}
\ee
The improper elements consist of
\be
P^{i} = \{i, 8 \tilde{S}_6, 3\tilde{\sigma}_h, 6 \sigma_d, 6 S_4 \}
\ee
where $\tilde{g}$ denotes a non-symmorphic operation, in which rotations or reflections are accompanied by translation along an appropriate fraction of a lattice vector.  $P^{0}$ is a proper subgroup of $Fd \bar{3}m$, while $P^{i}$ is generated by the product of the inversion operator (inversion is taken about one of the lattice sites) with the elements of $P^{0}$. The symmetry transformations, along
with the full action of the PSG, are outlined in Appendix \ref{PSG}.

The PSG of the monopole flux state has the following general structure, outlined in more detail in Appendix \ref{PSG}:\\
$\bullet$ {\bf Translations} : FCC translations, combined with the identity gauge transformation. \\
$\bullet${\bf $P^{0}$ space group elements} : These elements are symmetries when combined with appropriate gauge transformations, which induce a $\pi$ phase shift at some the sites in the unit cell. \\
$\bullet${\bf $P^{i}$ space group elements} : These elements are symmetries when combined with an appropriate gauge transformation, as above, and a time reversal transformation.\\
$\bullet${\bf Charge conjugation $C$} : The charge conjugation operator maps $c_i \rightarrow c^\dag_i$. \\

\subsection{ Restricting Perturbative Corrections Using PSG Invariance}

To deduce what restrictions PSG invariance imposes on the spectrum, we begin with a generic $4\times4$ quadratic Hamiltonian
\be
H^{(2)} = \sum_{ij} J_{ij} c^\dag_i c_j
\ee
The bonds $J_{ij}$ connect arbitrary sites in the lattice, but respect the lattice symmetries.  In what follows, we will use the PSG to restrict the possible quadratic terms, and show that all terms allowed by symmetry vanish at the Fermi surface.  Hence the Fermi surface of the monopole flux state is unaffected by PSG-preserving perturbations to the Hamiltonian. For simplicity we will drop the superscript $(2)$ in the remainder of
this section to simplify the notation.

Though the inversion $P$
and time reversal $T$ are broken in the mean field state, the combination $PT$ leaves both the full and mean field Hamiltonians invariant.  Terms invariant under this transformation have the form:
\be \label{J1}
(J' + i J'') c^\dag_x c_{x+\delta} + (J' -i J'') c^\dag_x c_{x-\delta}
\ee
and the Hamiltonian is real in momentum space.  Further, invariance under charge conjugation forces all spatial bonds to be purely imaginary: under $C$,
\ba \label{J2}
(J' + i J'') c^\dag_x c_{x+\delta} +(J' -i J'') c^\dag_{x+\delta} c_{x} \n
\rightarrow (-J' + i J'')c^\dag_{x} c_{x+\delta} +  (-J' - i J'')c^\dag_{x+\delta} c_x
\ea
so that $J' =0$ if $C$ symmetry is unbroken.
In momentum space, if we write the Hamiltonian as $\overline{\psi} H^{}(k) \psi$, Equations (\ref{J1}) and (\ref{J2}) imply that $H^{}(k)$ is real and an odd function of $k$.
We may express elements of the matrix $H^{}(\bk)$ as a superposition
\be \label{H2inkspace}
H^{}_{ab} (\bk) = \sum_\bR J_{\bR; a b} \sin\left[ \bk \cdot (\bR+\mathbf{r}_{ab}) \right]
\ee
where $\bR$ is an FCC lattice vector, the indices $a,b$ label sites within the unit cell, and $J_{\bR; ab}$ is the coupling between sites $a$ and $b$ separated by the lattice vector $\bR$, and the vector $ \mathbf{r}_{ab}$ in the unit cell.  This is the general form for a function periodic in the Brillouin zone.

\noindent
{\bf Diagonal Terms}
Let $H_{11} ... H_{44}$ be the diagonal elements of $H^{}$.  To restrict the form of $H_{11}$,  we consider the action
of all PSG operations that map site $1$ in the tetrahedral unit cell
onto itself.  These are (see Appendix \ref{PSG} for labels and actions
of the PSG elements) $\{C_1, C_1^2, \tilde{C}_{23}, \tilde{C}_{24},
\tilde{C}_{34} \}$, which transform $H_{11}$ in the following way:
\ba
H_{11}(k_x, k_y, k_z) &\stackrel{ C_1}{\rightarrow} & H_{11}(k_z, k_x, k_y) \n
&\stackrel{ C_1^2} {\rightarrow}& H_{11}(k_y, k_z, k_x) \n
&\stackrel{ \tilde{C}_{23}}{\rightarrow} & H_{11}(-k_x, -k_z, -k_y) \n
&\stackrel{ \tilde{C}_{24}}{\rightarrow} & H_{11}(-k_z, -k_y, -k_x) \n
&\stackrel{\tilde{C}_{34}}{\rightarrow} & H_{11}(-k_y, -k_x, -k_z)
\ea
which allows us to express $H_{11}$ in a form where its symmetries are manifest as:
\ba \label{h11_1}
H_{11}(k_x, k_y, k_z) &=& \frac{1}{6} [ H_{11}(k_x, k_y,k_z) + H_{11}(k_z, k_x, k_y)\n
&&+ H_{11}( k_y, k_z, k_x)  - H_{11}(k_x, k_z, k_y)\n
& & -H_{11}(k_z,k_y, k_x) -H_{11}(k_y, k_x, k_z) ]
\ea

Similarly we can relate $H_{22}, H_{33},$ and $H_{44}$ to $H_{11}$ by considering operations which interchange site 1 with sites $2, 3$, and $4$ respectively. These
imply:
\ba \label{h11_2}
H_{22 }(k_x, k_y, k_z) &=& H_{1 1}(k_y, k_x, -k_z) \n
H_{3 3}(k_x, k_y, k_z) &=& H_{1 1}(k_z, -k_y, k_x) \n
H_{4 4 }(k_x, k_y, k_z) &=& H_{1 1}(-k_x, k_z, k_y) \n
\ea
While multiple transformations map between each pair of diagonal elements, the group structure and invariance of $H_{11}$ under PSG transformations ensures that these mappings all yield the same result.

The reader should note that Eqs. (\ref{h11_1}) and (\ref{h11_2}) ensure that along the Fermi lines $\bk = \pm (1, \pm 1, \pm 1)$ all allowed diagonal terms vanish.

It is worth digressing to make one more comment on the diagonal terms. Using the symmetrized form of $H_{11}$ in Eq.~(\ref{h11_1}) above, we can rewrite the term in Eq.~(\ref{H2inkspace}) with a fixed $\bR$ and $a=b=1$ as
\ba \label{DiagTerms}
J_{\bR; 11} \sin(\bk \cdot \bR) &=& \frac{1}{6} J_{\bR; 11} [ \sin(k_x R_x +
k_y R_y+ k_z R_z) \n && + \sin(k_z R_x + k_x R_y+
k_y R_z ) \n &&+\sin( k_yR_x +k_zR_y +k_x R_z )\n &&
  - \sin(k_x R_x + k_z R_y+ k_y R_z) \n &&- \sin( k_z R_x+  k_y R_y+ k_x R_z)\n
  &&- \sin(k_y R_x + k_x R_y+ k_z R_z) ]
 \ea
\noindent
The form (\ref{DiagTerms})
vanishes if any two coefficients are equal; non-vanishing terms occur only for a sum of at
least three FCC translations.  Physically this corresponds to a
hopping between a site and its translate some three lattice vectors
distant.

\noindent
{\bf Off-Diagonal Terms}
As $H$ is real in momentum
space, $H_{ab} = H_{ba}$.  To restrict the form
of $H_{12}$, consider the action of all PSG elements which either
map sites $1$ and $2$ to themselves, or interchange them.  These
are $\{ \tilde{C}_{34}, \tilde{C}_{12}, C_z \}$, which transform
$H_{12}$ according to
 \ba
H_{12}(k_x, k_y, k_z) &\stackrel{\tilde{C}_{34}}{ \rightarrow} &- H_{12} (-k_y, -k_x, -k_z ) \n
&\stackrel{\tilde{C}_{12}}{ \rightarrow}& H_{21} (k_y, k_x, -k_z ) \n
&\stackrel{C_{z}}{ \rightarrow}& - H_{21} (-k_x, -k_y, k_z )
 \ea
Again, transformations mapping sites $1$ and $2$ onto other sites
in the unit cell can be used to deduce the form of the remaining
off-diagonal elements.  Hence
\ba
H_{12}(k_x, k_y, k_z) &\stackrel{C_{1}}{ \rightarrow} & H_{13}(k_z, k_x, k_y) \n
&\stackrel{C_{1}^2}{ \rightarrow} & H_{14}(k_y, k_z, k_x) \n
&\stackrel{C_{4}^2}{ \rightarrow} &- H_{23}(-k_y, k_z, -k_x) \n
&\stackrel{C_{2}}{ \rightarrow} &- H_{24}(-k_z, k_x, -k_y) \n
&\stackrel{C_y}{ \rightarrow} &- H_{34}(-k_x, k_y, -k_z)
\ea

This gives off-diagonal entries:
\begin{widetext}
 \be
 \begin{bmatrix}
0 &  H_{12}(k_x, k_y, k_z)& H_{12}(k_z, k_x, k_y) &H_{12}(k_y, k_z, k_x) \\
H_{12}(k_x, k_y, k_z) & 0 &  H_{12}(k_y, -k_z, k_x) & H_{12}( k_z, -k_x, k_y)\\
H_{12}(k_z, k_x, k_y)  &H_{12}(k_y, -k_z, k_x)  & 0 & H_{12}(k_x, -k_y, k_z)\\
H_{12}(k_y, k_z, k_x)  & H_{12}( k_z, -k_x, k_y) &H_{12}(k_x, -k_y, k_z) &0
\end{bmatrix}
 \ee
\end{widetext}
where again we can make the symmetries manifest by writing
\ba \label{h12symmetric}
H_{12}(k_x,k_y,k_z)& =& \frac{1}{4} [ H_{12}(k_x, k_y,k_z) + H_{12}(k_y, k_x, k_z )\\
&& + H_{12}(k_y, k_x, -k_z )
 + H_{12}(k_x, k_y,-k_z ) ] \ . \nonumber
\ea
Again, it is useful to focus on the contribution to $H_{12}$ from bonds with
a given $\bR$ which can now be seen to come with the factor:
 \ba \label{fsines}
\cos(k_z R_z) [ \sin( k_x R_x + k_y R_y+\frac{k_x+k_y}{2}) \n
 + \sin(k_y R_x + k_x R_y+\frac{k_x+k_y}{2})] \ .
 \ea
Eq. (\ref{fsines}) shows that $H_{12}(k_x,k_y,k_z)$ vanishes along the lines $(k, -k, -k), ( -k, k, -k), ( -k, -k, k)$. Of course, this can also be seen directly from Eq.~(\ref{h12symmetric}).

Now we may consider the fate of the monopole flux state's exotic Fermi
surface.  Since PSG rotations map between different Fermi lines,
it is sufficient to consider possible alterations to the spectrum
on Fermi line $(k,k,k)$.
The most general form that $H$ can have about the
line $(k,k,k)$ is:
 \be  \label{OddHs}
 H= \begin{bmatrix}
 0 & H_{12}(k,k,k) 	&
 H_{12}(k,k,k)& H_{12}(k,k,k) \\
 H_{12}(k,k,k) & 0 & 0 & 0 \\
 H_{12}(k,k,k) &0 & 0 & 0 \\
 H_{12}(k,k,k) & 0 & 0 &0
 \end{bmatrix}
 \ee
which has two zero eigenvalues. Thus terms allowed by symmetry add neither a
chemical potential nor a gap to any part of the Fermi lines, and
preserve the characteristic structure of the monopole flux state, with
2 low energy states about each Fermi line, and 4 low energy
states about the origin.

Note that nothing prevents the Fermi velocity $v_F$ from being
modified as a function of the momentum along the line.  Indeed,
(\ref{OddHs}) implies that the general form of $v_F$ is:
\begin{widetext}
 \ba
\frac{ v_F(k)}{\sqrt{2}} &=& \sum_{\bR} J_{\bR;1 2} (R_x+R_y+1/2) \cos(2k R_z) \cos 2k (R_y-R_x)
 +  \frac{J_{\bR; 1 1}}{3} ( R_x \sin (2k R_x ) \sin 2k (R_z  -R_y ) \n
&& + R_y \sin (2k R_y ) \sin 2k ( R_x -R_z)
    + R_z \sin (2kR_z ) \sin 2k(R_y -R_x )   ) \ .
 \ea
\end{widetext}

\subsection{Time reversal and Parity}

One striking feature of $H_{MF}$ is that it is {\em odd} under both $T$
and $P$, reminiscent of the chiral spin state first described in
Ref.~\onlinecite{WWZ}.  Though $T$ is naively broken, some care must be taken to show that the apparent $T$ breaking is physical and that $|\psi \rangle, T| \psi \rangle $ are gauge inequivalent states\cite{HermeleMother}.  Readers familiar with this subtlety from discussions of $T$ breaking on the square lattice, should note that the pyrochlore lattice is not bipartite and hence naive time reversal is no longer equivalent to particle-hole conjugation. But most directly, as explained in Ref.~\onlinecite{WWZ}, the operator
\be
E_{ijk} =S_i \cdot (S_j \times S_k)
\ee
where the spins $i$, $j$, and $k$ lie in a triangular plaquette, is odd under
$T$ and $P$. Hence if $\langle E_{ijk} \rangle_{|\psi\rangle} \neq 0$, the state
$  |\psi\rangle$ breaks time reversal.

At mean field level,
\be
\frac{-i}{2}\langle E_{123} \rangle = \langle \chi_{12} \chi_{23} \chi_{31} \rangle  -
 \langle \chi_{13} \chi_{32} \chi_{21} \rangle
\ee
and states with an imaginary flux through triangular plaquettes are $T$-breaking.
For the monopole flux state, we have confirmed numerically that this $T$-breaking
is robust to Gutzwiller projection; the results are shown in Table \ref{PTTab}.

\begin{table}[ht]
\begin{center}
\begin{tabular}{|c|c|}
\hline
Lattice size & $\langle E_{\triangle} \rangle$ \\
\hline
$3\times 3\times 3$ &  $ 0.039$ \\
$5\times 5 \times 5$ & $0.043 $\\
\hline
\end{tabular}
\end{center}
\caption{Expectation values of the $T$-breaking operator $\langle E_{\triangle} \rangle$
for triangular faces of the tetrahedra.} \label{PTTab}
\end{table}

We also note the curiosity that at infinite $N$, the spectrum-preserving nature of $T$ and $P$
allows us to construct additional symmetries which are \emph{not},
however, symmetries of the full $H$.
Particle-hole symmetry at each $k$ allows us to construct
the following 2 discrete symmetries of $H_{MF}$
:
 \ba
\tilde{T}:  |\psi(x, t) \rangle \rightarrow \alpha^{(0)} |\psi (x, -t) \rangle \n
\tilde{ P}:  |\psi (x,t)\rangle \rightarrow \alpha^{(0)} |\psi (-x, t) \rangle \n
 \ea
where $\alpha^0$ was defined in Eq.~(\ref{alpha0}). Both of these commute with the non-interacting Hamiltonian: since
$\alpha_0^{-1} H \alpha_0 = -H$, we have
 \ba
\langle \psi| \tilde{T}^{\dag} H \tilde{T} | \psi\rangle &= & - \langle \psi| H^*
|\psi \rangle \n
  & =& \langle \psi | H | \psi \rangle \n
\langle \psi| \tilde{P}^{\dag} H \tilde{P} | \psi \rangle &= & - \langle \psi| H^T |\psi \rangle \n
 & =&  \langle \psi | H | \psi\rangle
 \ea
 The matrix structure of $\alpha_0$ is such that $\tilde{T}$ and
$\tilde{P}$ are not symmetries of the full Hamiltonian, however,
and will not be robust to perturbative corrections about mean field.

\subsection{PSG symmetry and perturbation theory in the long wavelength
  limit}

We have established that invariance under the PSG transformations and
charge conjugation forbid both mass and chemical potential terms on the Fermi lines.  Here we explore how these PSG symmetries are realized as symmetries of the linearized low-energy theory away from the origin, and hence see in that setting why they are protected perturbatively.

Consider the linearized theory about the Fermi line $\bm{l_1}=(k,k,k)$.  A
general Hamiltonian in the $2\times 2$ space of low-energy states can
be expressed as:
\begin{eqnarray} \label{Hlo}
H(\bk)  &=& \psi^\dag_{1}(\bk)  h(\bk) \psi_1 (\bk) \\
h(\bk) &=&  \mu(k)  + m(k) \sigma_3  +  \varepsilon(k, v) (\cos(\theta) \sigma_1 +
\sin(\theta) \sigma_2)  \nonumber
\end{eqnarray}
where $k$ is the component of the momentum $ \bk$ along the line, and $(v,
\theta)$ are the magnitude and angle respectively of the momentum
perpendicular to the line.  Here
\begin{eqnarray} \label{psis}
\p_{11} (\bk) =(0, \w^2, 1, \w)\cdot\bm{c_\alpha} (\bk) \nonumber \\
\p_{12} (\bk) = (0, \w, 1, \w^2)\cdot\bm{c_\alpha}(\bk)
\end{eqnarray}
with $\w = e^{2 \pi i/3}$.  The states (\ref{psis}) are
eigenstates of the rotation operator $C_1$ which rotates about
the $(1,1,1)$ direction, with $C_1 \p _{1j} =\w^j \p_{1j}$.

Under charge conjugation,
\ba
\p_{11} (\bk) \rightarrow \p^\dag_{12} (-\bk) \n
\p_{12} (\bk) \rightarrow \p^\dag_{11} (- \bk)
\ea
The corresponding symmetry operator in the continuum theory is
\be \label{CTrans}
C : \psi_1 (v, \theta) \rightarrow \sigma_1 [\psi_1^\dag (v, \pi - \theta )]^T
\ee
with the Fermi surface points  at $k$ and $-k$ interchanged.
This implies $m(-k) = m(k)$, and $\mu(-k) = - \mu(k)$.

Further, an analysis of the PSG transformations reveals that the glide
rotations ($\tilde{C}_{ij}$) map clockwise rotating states to
counter-clockwise states while reversing the direction of
the corresponding Fermi line:
 $\psi_{i1} (k l_i)  \rightarrow \psi_{i2} (-k l_i)$.  This
 transformation leaves the mean field Hamiltonian invariant.  In the $2 \times
 2$ basis, this is because
 \be \label{C34}
T: |\psi(v, \theta)\rangle  \rightarrow \sigma_1 |\psi(v,\pi -\theta)\rangle
 \ee
is a symmetry of the mean field Hamiltonian.  Note that the momentum
transformation can be realized in 3 dimensions by a $\pi$
rotation about the line $x=y$, and hence should also send $k
\rightarrow -k$, though there is no way to deduce this from the
form of the mean field eigenstates. The symmetry transformation
(\ref{C34}) reverses the sign of the mass term, but not of the
chemical potential, implying that $m(-k) = - m (k)$ and $\mu(k)
= \mu(-k)$.  Hence we conclude that in the continuum theory about a
given Fermi line, the symmetries $C$ and $T$ prevent a gap or chemical
potential from arising.

One might ask why we have not considered mass gaps of the form $m
\sigma_1$ or $m \sigma_2$; both of these choices turn out to
violate either (\ref{CTrans}) or (\ref{C34}).  Indeed, both
choices explicitly break the rotational symmetry of the spectrum
about the Fermi line.

\section{Concluding Remarks} \label{Conc}

We have discussed an interesting mean field (large $N$) solution to the
Heisenberg model on the pyrochlore lattice. This is a P and T breaking state in which all triangular
plaquettes have an outward flux of $\pi/2$.  After Gutzwiller projection,
this state has lower energy than all other mean field states considered,
including the simplest dimerized state.  Its low-energy physics is rather
striking, with a spinon Fermi surface of lines of nodes preserving the discrete
rotational symmetries of the lattice.  The symmetries of the Hamiltonian
suggest that this Fermi surface is perturbatively stable and thus should characterize a stable spin liquid phase, at least at sufficiently large $N$.

However, our analysis of stability is thus far based on only on symmetries and does not rule out dynamical instabilities. The study of such instabilities requires adding back in the gauge fluctuations that are suppressed at $N = \infty$ and studying the coupled system consisting of spinons and gauge fields. In the well studied case of two dimensional algebraic spin liquids it took a while to understand that this coupled system could, in fact, support a gapless phase at sufficiently large $N$ despite the compactness of the gauge fields. In the present problem there is also the specific feature that at $N=2$, as the background flux per plaquette is $U(1)$, such an analysis should incorporate fluctuations of an $SU(2)$ gauge field \cite{WenPSG}.We defer addressing this set of questions to future work \cite{wip}.

Finally, we note that our initial motivation in this study was to see if we could construct a fully symmetric spin liquid on the pyrochlore lattice for $S=1/2$ in contradiction with previous studies using other techniques. We have not succeeded in that goal and, as the technique in this paper has produced a pattern of symmetry breaking distinct from the ones considered previously, the fate of the $S=1/2$ nearest-neighbor Heisenberg antiferromagnet on the pyrochlore lattice remains undeciphered.

\begin{acknowledgements}
We would like to thank Eduardo Fradkin, Roderich Moessner, Ashvin Vishwanath and Michael Hermele for instructive discussions. We would also like to acknowledge support from NSF Grant No. DMR 0213706. F. J. B. acknowledges the support of NSERC.
\end{acknowledgements}

\appendix
\section{\label{Gutz} Monte Carlo and Projected Wavefunctions}

Gutzwiller projection can be carried
out exactly for finite systems using the projector in the Slater determinant
basis, using the method of \onlinecite{gros}.
At half filling, the Slater
determinant is represented as a product of a spin up and a  spin down
determinant of equal size ($N/2$) where $N$ is the number of sites in the
lattice. Each site is represented exactly once in either the spin up or the
spin down matrix, yielding a wavefunction which obeys the
single occupation constraint on all lattice sites and is a total spin singlet.
The problem then reduces to the evaluation of expectation values of operators
for wavefunctions of finite systems with definite spin distributions on the
lattice:
\begin{equation}
\langle O\rangle=\frac{\langle\psi|O|\psi\rangle}{\langle\psi|\psi\rangle},
\end{equation}
 where $|\psi\rangle=\sum_\alpha\langle\alpha|\psi\rangle|\alpha\rangle$, and $|\alpha\rangle$ is a specific distribution of spins on the lattice,
\begin{equation}
|\alpha\rangle=\prod_i c^\dagger_{R_i,\uparrow}\prod_j c^\dagger_{R_j,\downarrow}|0\rangle.
\end{equation}
The expectation value of the operator $O$ is evaluated by summing
over all spin configurations on the lattice.  To evaluate this
sum we follow the approach of  \onlinecite{gros}, which we will review
here.  The expectation value is given by:
\begin{equation}
 \langle O\rangle=\sum_\alpha\Bigl(\sum_\beta\frac{\langle\alpha|O|\beta\rangle\langle\beta|\psi\rangle}{\langle\alpha|\psi\rangle}\Bigr )\frac{|\langle\alpha|\psi\rangle|^2}{\langle\psi|\psi\rangle}
=\sum_\alpha f(\alpha)\rho (\alpha),
\end{equation}
where
\begin{subequations}
\begin{align}
f(\alpha)&=\sum_\alpha\Bigl(\sum_\beta\frac{\langle\alpha|O|\beta\rangle\langle\beta|\psi\rangle}{\langle\alpha|\psi\rangle}\Bigr ),\\
\rho(\alpha)&=\frac{|\langle\alpha|\psi\rangle|^2}{\langle\psi|\psi\rangle}.
\end{align}
\end{subequations}
Here $\rho(\alpha)\geq0$ and $\sum_\alpha\rho(\alpha)=1$, which makes $\rho(\alpha)$ a probability distribution.
The expectation value can be rewritten to resemble a weighted sample using $\rho(\alpha)$ evaluated using a Monte Carlo sampling. The evaluation is executed using a random walk in configuration space with the weight $\rho(\alpha)$. The transition probability $T_{\alpha\alpha'}$ of the Monte Carlo step is
\begin{equation}
T_{\alpha\alpha'}=\begin{cases}1& \rho(\alpha')>\rho(\alpha),\\
\rho(\alpha')/\rho(\alpha)&\rho(\alpha')<\rho(\alpha).\end{cases}
\end{equation}
The configuration $\alpha'$ is generated by exchanging a randomly selected  pair of oppositely oriented spins.   We also calculate various operators in the mean field wavefunction to test the accuracy of the algorithm. In this case the one particle per site constraint is not imposed; the configuration $\alpha'$ is generated by moving an up or down electron at random  to another empty up or down site.

Pyrochlore has an FCC lattice with a four point basis. We use the
rhombohedral unit cell for  the Monte Carlo evaluation, using
boundary conditions periodic along the fcc directions.
 The spin correlations turn out to be quite insensitive to  boundary conditions for the lattice sizes that we have considered.

To compute the Slater determinant, the mean field Hamiltonian (\ref{hmonopole}) is diagonalized in
the band eigenstates
\begin{equation}
H=\sum_{a,b,\bm{k}}c^{\dagger\alpha}_{a\bm{k}}H_{ab}(\bm{k})c_{b\bm{k}\alpha}=\sum_{\nu\bm{k}}E_\nu(\bm{k})
a^{\dagger\alpha}_{\nu\bm{k}}a_{\nu\bm{k}\alpha},
\end{equation}
where  $c^{\dagger\alpha}_{a\bm{k}}$ is the Fourier transform
\begin{equation}
c^{\dagger\alpha}_{a\bm{k}}=\sum_{\bm{k}}e^{i\bm{k}\cdot(\bm{R}_i +\bm{r}_{a_i}/2)}c^{\dagger\alpha}_{\bm{R}_i +\bm{r}_{a_i}/2}.
\end{equation}
Here $\bm{r}_{a_i}/2$ refer to the points in the four site basis
of the tetrahedral unit cell. If we assume that the bands
$\nu=1,2$ are filled,  the operator $c^{\dagger\alpha}_{a\bm{k}}$
can be expressed as
\begin{equation}
c^{\dagger\alpha}_{a\bm{k}}=S_{a,1}(\bm{k})a^{\dagger\alpha}_{1,\bm{k}} +S_{a,2}(\bm{k})a^{\dagger\alpha}_{2,\bm{k}},
\end{equation}
where  the matrix $S^\dagger_{a,\nu}(\bm{k})$ diagonalizes the Hamiltonian.  The mean field ground state is just the Fermi sea filled to the appropriate Fermi level which, in this case, is the 1st Brillouin zone boundary.
\begin{equation}
|\Phi\rangle_{mean}=\prod_{\bm{k}<\bm{k}_f,\alpha}a^{\dagger\alpha}_{1,\bm{k}}a^{\dagger\alpha}_{2,\bm{k}}|0\rangle
\end{equation}
We  rewrite  the mean field Fermi sea  $|\Phi\rangle_{mean}$ in first quantized form;
\begin{equation}
\Phi=\sum_{U,D}\Phi_{U\uparrow}\Phi_{D\downarrow}|U\rangle |D\rangle,
\end{equation}
where
\begin{align}
|U\rangle&=c^\dagger_{\bm{R_{1,a_1}}\uparrow}\cdots c^\dagger_{\bm{R_{N_L/2,a_{N_L/2}}}\uparrow}|0\rangle,\\
|D\rangle&=c^\dagger_{\bm{R'_{1,a_1}} \downarrow}\cdots c^\dagger_{\bm{R'_{N_L/2,a_{N_L/2}}}\downarrow}|0\rangle.
\end{align}
The basic Slater determinant wavefunction for each spin orientation  is of the form
\begin{equation}
\Phi(\bm{R_{i,a_i}},\bm{k_j},\nu)_\alpha=\text{Det} [\phi(\bm{R_{i,a_i}}:\bm{k_j},\nu)]_\alpha,
\end{equation}
where $\phi(\bm{R_{i,a_i}}:\bm{k_j},\nu)$ is the single particle wavefunction of the electron at $\bm{R_{i,a_i}}=\bm{R}_i +\bm{r}_{a_i}/2$. The  wave number $\bm{k_j}$ refers to a point in the conjugate lattice  in the rhombohedral Brillouin zone, and  $\nu$ is the band index. In terms of the matrix $S_{a,\nu}(\bm{k})$, the single particle wavefunction is
\begin{equation}
\phi(\bm{R_{i,a_i}}:\bm{k_j},\nu)=S_{a,\nu}(\bm{k_j})e^{i\bm{k_j}\cdot(\bm{R}_i +\bm{r}_{a_i}/2)}.
\end{equation}
In the above, $U=\{\bm{R_{1,a_1}},\cdots, \bm{R_{N_L/2,a_{N_L/2}}}\}$ is the set of lattice sites occupied by up spin electrons, and   $D=\{\bm{R'_{1,a_1}},\cdots, \bm{R'_{N_L/2,a_{N_L/2}}}\}$ is the set of lattice sites occupied by down spin electrons.

Gutzwiller projection is imposed by ensuring that the two sets $U$ and $D$ have no elements in common. The Monte Carlo update is performed by exchanging rows selected at random from the $\Phi_{U\uparrow}$ and $\Phi_{D\downarrow}$ matrices.

 The calculation of the transition probability $T_{\alpha\alpha'}$  involves the calculation of  determinants of matrices of size $N_L/2\times N_L/2$, an $O(N_L^3)$ operation. The  algorithm of Ceperley, Chester  and Kalos \cite{cck} reduces this to an $O(N_L^2)$ operation for the special case of updates involving one row or column. The matrix $M^U=\Phi_{U\uparrow}$ and the transpose of its inverse
$\overline{M}^U$ (similarly for $\Phi_{D\downarrow}$) are stored at the beginning of the Monte Carlo evolution. If the update changes the $a$th row $M^U_{aj}\rightarrow A_j, M^U\rightarrow M'^U$, the transition matrix is $T_{\alpha\alpha'}=\text{Det}M'^U/\text{Det}M^U$:
\begin{equation}
\frac{\text{Det}M'^U}{\text{Det}M^U}=\sum_{j=1}^{N_L/2} A_j\overline{M}^U_{ja} =r.
\end{equation}
This is due to the fact that $\text{Det}(M^U\overline{M}^U)$ is the matrix of cofactors. If the move is accepted, the inverse matrix can be updated using $O(N_L^2)$ operations:
\begin{equation}
\overline{M}^U_{ji}=\begin{cases}\overline{M}^U_{ji}/r & i=a,\\
    \overline{M}^U_{ji} -\overline{M}^U_{ja}\sum_{k=1}^{N_L/2}A_k\overline{M}^U_{kj}/r & i\neq a.
             \end{cases}
\end{equation}

The evaluation of operator expectations values has to be handled with care as we are dealing with fermions. The relative sign of determinants must be tracked in a consistent way.
Thus we write all spin configurations in the order:
\begin{equation}
|\alpha\rangle =c^\dagger_{\bm{R_{1,a_1}}\uparrow}\cdots c^\dagger_{\bm{R_{N_L/2,a_{N_L/2}}}\uparrow},c^\dagger_{\bm{R'_{1,a_1}} \downarrow}\cdots c^\dagger_{\bm{R'_{N_L/2,a_{N_L/2}}}\downarrow}|0\rangle.
\end{equation}

The wavefunction is given by $\langle\Phi|\alpha\rangle =\Phi_{U\uparrow}\Phi_{D\downarrow}(\alpha)$, the Slater determinant
 eigenfunction with the given spin distribution $\alpha$. We are interested in the operators $\sum S_{\bm{R_{i,a_i}}}^z\cdot S_{\bm{R_{j,a_j}}}^z$ and $\sum (S_{\bm{R_{i,a_i}}}^+\cdot S_{\bm{R_{j,a_j}}}^- +S_{\bm{R_{i,a_i}}}^-\cdot S_{\bm{R_{j,a_j}}}^+)$. To keep track of the proper sign specification we  express the spin operators in terms of fermionic operators with the same order as the spin configuration:
\begin{align}
S_{\bm{R_{i,a_i}}}^z\cdot S_{\bm{R_{j,a_j}}}^z&=(n_{\bm{R_{i,a_i}}\uparrow}-n_{\bm{R_{i,a_i}}\downarrow})(n_{\bm{R_{j,a_j}}\uparrow}-n_{\bm{R_{j,a_j}}\downarrow})/4,\\
S_{\bm{R_{i,a_i}}}^+\cdot
S_{\bm{R_{j,a_j}}}^-&=c^\dagger_{\bm{R_{i,a_i}}\uparrow}c_{\bm{R_{i,a_i}}\downarrow}c^\dagger_{\bm{R_{j,a_j}}\uparrow}c_{\bm{R_{j,a_j}}\downarrow}
\\
&=-c^\dagger_{\bm{R_{i,a_i}}\uparrow}c^\dagger_{\bm{R_{j,a_j}}\uparrow}c_{\bm{R_{i,a_i}}\downarrow}c_{\bm{R_{j,a_j}}\downarrow}.
\end{align}
The  amplitude $S_{\bm{R_{i,a_i}}}^+\cdot S_{\bm{R_{j,a_j}}}^-|\alpha\rangle$ is the  determinant wavefunction $\Phi_{U\uparrow}\Phi_{D\downarrow}(\alpha')$ with the rows of $\Phi_{U\uparrow}$ and $\Phi_{D\downarrow}$ changed as described above. $S_{\bm{R_{i,a_i}}}^z\cdot S_{\bm{R_{j,a_j}}}^z|\alpha\rangle$ is easier to evaluate as it is diagonal.

As a result of the four site basis there are only $N_L/4$ lattice points in the Brillouin zone. Therefore, the  accuracy of the Monte Carlo evaluation is limited by the number of points in $\bm{k}$ space. We have used a lattices of size
$5\times5\times5$ (500 sites).

\begin{figure}[th]
\begin{center}
\epsfxsize=3.in
\epsffile{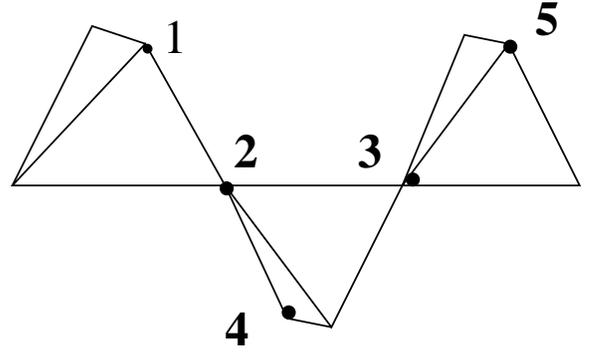}
\caption[Sites for Spin Correlation Functions.]{Sites for the spin correlation functions calculated in  Table \ref{tab:table1}.}
\label{kagome1}
\end{center}
\end{figure}

A basic Monte Carlo `move' consists of $2N_L$ updates followed by a sampling.  The first 10000 moves were used for thermal relaxation and were discarded. $50000$ samples were used for the evaluation of expectation values.  These were divided into $10$ sets and the average in each set was used to estimate the statistical fluctuations and error bars of the Monte Carlo evaluation.
  To check the accuracy of the algorithm  and the effect of finite lattice size  we  evaluated spin correlation functions of the mean field states and compared with results from the numerical evaluation of Green's functions. The Monte Carlo results are quite close to the expected values (see  Table \ref{tab:table1}). The site indices of the spins refer to Fig. \ref{kagome1}.

\begin{table}
\caption{Spin Correlations from Variational Monte Carlo.  To check the accuracy of the algorithm, correlators are also calculated at mean-field, using both Monte Carlo (MC) and Green's function (G) approaches.  }
\label{tab:table1}
\begin{center}
\begin{tabular}{|l|c|c|c|c|}
\hline
Trial Wavefn.& $S^z_1 S^z_2$& $S^z_1 S^z_3$&$S^z_1 S^z_4$& $S^z_1 S^z_5$\\
{\textbf{mean field}}& & & & \\
Flux (G)& --0.01388&--0.00188&--0.00097&--0.00097\\
Flux (MC)&--0.01386&--0.00192&--0.00103&--0.00084\\
    & $\pm$0.00003&$\pm$0.00004&$\pm$0.00005&$\pm$0.00006\\
Monopole (G)&--0.01745&0.00000&0.00000&--0.00087\\
Monopole (MC)&--0.01713&--0.00002&--0.00008&--0.00085\\
    &$\pm$0.00004&$\pm$0.00001&$\pm$0.00002&$\pm$0.00006\\
{\textbf{Projected}}&&&&\\
Flux &--0.04169&--0.00137&0.0029&--0.00270\\
    &$\pm$0.00004&$\pm$0.00005&$\pm$0.0001&$\pm$0.00008\\
Monopole &--0.0497&0.00631&0.00528&--0.00499\\
    &$\pm$0.0001&$\pm$0.00002&$\pm$0.00004&$\pm$0.00005\\
    \hline
\end{tabular}
\end{center}
\end{table}

\section{ The PSG of the monopole flux state} \label{PSG}

\subsection{Symmetries of the Pyrochlore Lattice}

The space group $Fd\bar{3}m$ of the pyrochlore lattice
consists of  the 24 element tetrahedral point group $\bar{4} 3 m$, and a further  24 non-symmorphic elements.  We will briefly describe
the actions of these symmetry operations here.  All vectors are expressed in the
basis of the standard cubic FCC unit cell.

The actions of the tetrahedral point group  fix the position of one
tetrahedron's center (at e.g. $(a/8, a/8, a/8)$).  Its elements are:
\begin{enumerate}
\item The identity
\item $8C_3$ : there are four 3-fold axes, one passing through each vertex and the
center of the opposite face.  Rotations about this axis permute the 3 vertices
not on the axis.  These we label $C_1.. C_4$, $C_1^2 .. C_4 ^2$, where $C_i, C_i^2$ fix site $i$ of the tetrahedral unit cell.
\item $ 3 C_2$: There are 3 2-fold axes, parallel to
the $x, y$, and $z$ axes. Each axis bisects a pair of edges on the
tetrahedron; the ensuing rotation exchanges pairs of vertices.  These we label $C_x, C_y, C_z$.
\item $6 \sigma_d$: A plane of reflection passes through each edge, and out
the center of the opposite face.  These planes lie on the diagonals with respect
to the FCC cubic unit cell, and hence are called diagonal reflections.
\item $6 S_4$: An improper rotation of degree $4$ about the axis bisecting 2
edges (parallel to the $x, y,$ or $z$ axis) is also a symmetry.  The
tetrahedron is rotated by $ \pi /2$ about e.g. $(1/8, 1/8, z)$ and
reflected through the plane $z=1/8$.
This operation squared produces one of the 2-fold rotations, so each axis
contributes 2 group elements.
\end{enumerate}

The remaining non-symmorphic elements, which we distinguish from their symmorphic counterparts using the notation $\tilde{S}$, are:
\begin{enumerate}
\item $6 \tilde{C}_4$: There are three 4-fold screw axes: $(3/8, 1/8, z)$,
  $(3/8, y, 1/8$, and $(x, 3/8, 1/8)$.    The symmetry rotates the
lattice by $\pi/4$ about such an axis, and translates by $1/4$ of the
side length of the FCC cubic unit cell along the axis.  Each
axis accounts for 2 elements of the quotient group, as $C_4^2 = t C_2$, with
$t$ an FCC translation and $C_2$ one of the 2-fold rotations of the
  point group.  These we label $\tilde{C}_x,\tilde{C}_x^2, \tilde{C}_y, \tilde{C}_y^2,
\tilde{C}_z, \tilde{C}_z^2$.
\item $ 6 \tilde{C}_2$: Along each of the 6 edges of the tetrahedron (the FCC basis
vectors) there is a 2-fold screw axis.  The lattice is translated along
the edge of a tetrahedron, then rotated by $\pi$ about this edge. These we label $\tilde{C}_{ij}$, where $\tilde{C}_{ij}$ has a screw axis along the line joining sites $i$ and $j$.
\item $3 \tilde{\sigma}_h$: The $x,y$ and $z$ planes of the cubic unit cell each
contain a horizontal glide plane.  The lattice is translated along an
FCC vector in the plane, e.g. by
$ (1/4, 1/4, 0)$, and then reflected through the plane -- in our
example, through $z=0$.
\item $i$ The lattice is inverted about the origin.
\item $8 \tilde{S}_6$: The products of the $8$ $C_3$ rotations with the inversion
give 8 improper 3-fold rotations.  (These are not in the point group because
they map a single tetrahedron onto its neighbor.)
\end{enumerate}

For our purposes these 48 elements divide into $24$ $C$ elements involving
pure rotations and translations, and $24$ $S$ elements involving improper
rotations, reflections, or inversions.  The $S$ elements are not
symmetries of the monopole flux state, as they map monopoles to
anti-monopoles; to construct the appropriate symmetry elements they
must be combined with a time reversal transformation.  Since all such
elements can be expressed as a product of a $C $ element with the
inversion, this is simply a consequence of the fact that while P and T
are separately broken in the monopole flux state, the combination PT is
still a symmetry.

\subsection{The PSG}

Here we list the PSG transformation rules for the symmetry
operations described above.  Throughout, we use the gauge
illustrated in Fig. \ref{monopolefig}, in which all bonds are
either ingoing or outgoing from site $`1'$; starting from a
different gauge will permute the gauge transformations listed
here (note that this has no effect on which bonds are allowed or
disallowed by the PSG, however).  Note that these tables only
show the mapping between the site labels $1 ...4$; it is
important to bear in mind the effect of the translations in the
case of the non-symmorphic elements, which reverse the directions
between sites by interchanging up and down triangles.  To this
end we also include a table of momentum transformations under
these operations.

Since $S$ elements are products of $C$ elements and inversion, it
is sufficient to consider the 24 rotation operations, together
with the operator $i T$.  Note that group multiplication in the
PSG is valid modulo the global gauge transformation $c_i
\rightarrow - c_i$, which clearly does not alter the
Hamiltonian.  Thus only the relative phases of the $4$ sites in
the tetrahedral unit cell are relevant to the PSG transformation.

\begin{table}[htb]\label{tab:tab3}
\begin{center}
\caption[PSG of Spinons under Proper Rotations]{Action of PSG Point Group Rotations on spinon operators.}
\begin{tabular}{|c|c|c|c|c|c|c|c|c|c|c|c|}
\hline
$1$&$C_1$&$C_1^2$&$C_2$&$C_2^2$&$C_3$&$C_3^2$&$C_4$&$C_4^2$&$C_x$&$C_y$&$C_z$\\\hline\hline
$c_1$&$c_1$&$c_1$&$ c_4$&$- c_3$ &$ c_2$&$- c_4$ &$ c_3$&$- c_2$&
$- c_4$ &$ -c_3$&$ c_2$\\
$c_2$&$c_3$&$c_4$&$- c_2$&$- c_2$&$ c_4$&$ c_1$&$- c_1$&$ c_3$&
$ -c_3$ &$ c_4$&$- c_1$\\
$c_3$&$c_4$&$c_2$&$- c_1$&$ c_4$&$- c_3$&$- c_3$&$ c_2$&$ c_1$&
$ c_2$ &$ c_1$&$ c_4$\\
$c_4$&$c_2$&$c_3$&$ c_3$&$ c_1$&$- c_1$&$ c_2$&$- c_4$&$- c_4$&
$ c_1$ &$- c_2$&$ -c_3$\\
 \hline
  \hline
 $k_x $ & $ k_z$ & $ k_y$ & $- k_z$ & $ k_y$ & $
k_z$ & $ -k_y$ & $- k_z$ & $ -k_y$ &
$ k_x$ & $- k_x$ & $ -k_x$ \\
$k_y$ & $ k_x$ & $ k_z$ & $ k_x$ & $- k_z$ & $ -k_x$ & $ -k_z$ & $ -k_x$ & $ k_z$ &
$ -k_y$ &$ k_y$ &$- k_y$ \\
$k_z$ &$ k_y$ & $ k_x$ & $ -k_y$ & $- k_x$ & $ -k_y$ & $ k_x$ & $ k_y$ & $ -k_x$ &
$- k_z$ & $- k_z$ & $ k_z$ \\
\hline
\end{tabular}
\end{center}
\end{table}

\begin{table}
\begin{center}
\caption{PSG action of screw rotations}
\begin{tabular}{|c|c|c|c|c|c|c|c|c|c|c|c|c|}
\hline
\hline
 & $\tilde{C}_{12}$ & $\tilde{C}_{13}$ & $\tilde{C}_{14}$ & $\tilde{C}_{23}$ & $\tilde{C}_{24}$ & $\tilde{C}_{34}$ & $\tilde{C}_x$ &$\tilde{C}_y$ &$\tilde{C}_z$ & $\tilde{C}_x^3$ &$\tilde{C}_y^3$ &$\tilde{C}_z^3$ \\
\hline
$c_1$ & $c_2$ & $c_3$ & $c_4$ & $-c_1 $&$ -c_1$& $-c_1$ &
$c_2$ & $c_4$ & $c_3$ & $c_3$ & $c_2 $&$ c_4$\\
$c_2$ & $c_1$ &$ -c_2$ & $c_2$ &$ c_3$ &$ c_4$&$ c_2$ &
$-c_4$ & $c_1$ & $c_4$ & $c_1$ & $c_3 $&$ -c_3$\\
$c_3$ & $c_3$& $ c_1$&  $ -c_3$& $ c_2$&$ c_3$&$c_4$ &
$c_1$ & $c_2$ & $-c_2$ & $c_4$ & $-c_4 $&$ c_1$\\
$c_4$ & $-c_4$& $c_4$ & $ c_1$& $ c_4$&$  c_2$&$c_3$ &
$c_3$ & $-c_3$ & $c_1$ & $-c_2$ & $c_1 $&$ c_2$\\
\hline
\hline
$k_x $ & $ k_y$ & $ k_z$ & $- k_x$ & $ -k_x$ & $ -k_z$ & $ -k_y$
& $k_x$ & $ k_x$ & $ k_z$ & $- k_z$ & $ -k_y$ & $ k_y$ \\
$k_y$ & $ k_x$ & $ -k_y$ & $ k_z$ & $- k_z$ & $ -k_y$ & $ -k_x$
& $ -k_z$ & $ k_z$ & $ k_y$ &$ k_y$ &$k_x$ & $ -k_x$ \\
$k_z$ &$ -k_z$ & $ k_x$ & $ k_y$ & $- k_y$ & $ -k_x$ & $ -k_z$
& $ k_y$ & $ -k_y$ & $- k_x$ & $ k_x$ & $ k_z$ & $ k_z$ \\
\hline
\end{tabular}
\end{center}
\end{table}

\subsection{Action of the PSG on Low-Energy Eigenstates}

Recall that the eigenstates of the low-energy excitations about the
Fermi surface can be expressed in terms of eigenstates of the point
group rotation operators $8C_3$ (c.f. (\ref{psis})).

Tables \ref{Transforms} and \ref{MoreTransforms} list the action
of the rotation elements of the PSG on these low-energy states.
Since (\ref{psis}) is PT invariant, it is sufficient to consider
the action of the rotation subgroup; the other PSG elements are
combinations of rotations with the inversion, and cannot procure
new information about the low-energy behavior.   Note that
proper rotations always map clockwise rotating states to
clockwise states, and counter-clockwise to counter-clockwise. The
improper rotations, conversely, map counter-clockwise rotating
states into clockwise rotating states, and vice versa.

\begin{table} \label{Transforms}
\caption{Effect of Point Group rotations on low-energy eigenstates.}
\begin{tabular}{|c|c|c|c|c|c|c|c|}
\hline
\hline
 & $C_1$ & $C_2$ & $C_3$ & $C_4$ & $C_x$ & $C_y$& $C_z$ \\
\hline
$\p_{11}$ & $\w \p_{11}$ & $\w^2 \p_{31}$ & $-\w \p_{41}$ & $-\w^2 \p_{21} $&
 $- \p_{41}$ & $ \w \p_{31}$ & $ \w \p_{21}$\\
$\p_{12}$ & $\w^2 \p_{12}$ &$ \w \p_{32}$ & $-\w^2 \p_{42}$ &$ -\w \p_{22}$ &
$ -\p_{42} $&$ \w^2 \p_{32}$& $\w^2 \p_{22} $\\
$\p_{21}$ & $\p_{41} $& $\w \p_{21} $&  $\p_{11} $& $\w \p_{31}  $ &
$ -\p_{31}$&$ -\w^2 \p_{41} $& $-\w^2 \p_{11} $\\
$\p_{22}$ & $\p_{42} $& $\w^2 \p_{22}$ & $\p_{12} $& $\w^2 \p_{32} $&
$ - \p_{32}$&$  -\w \p_{41}$ & $ -\w \p_{12} $\\
$\p_{31}$ & $-\w \p_{21}$ & $\p_{41} $& $\w \p_{31}$ & $-\p_{11} $
&$ \p_{21} $&$ -\w^2 \p_{11}$& $ \w^2 \p_{41}$\\
$\p_{32}$ & $-\w^2\p_{22}$ & $\p_{42} $&$ \w^2 \p_{32}$ & $- \p_{12}$
&$ \p_{22} $&$ -\w \p_{12}$& $ \w \p_{42} $\\
$\p_{41}$ & $-\w^2 \p_{31}$ & $\w \p_{11}$ & $- \w^2 \p_{21}$ &$  \w \p_{41}$
&$ \p_{11}$&$ \w \p_{21}$ & $ -\w \p_{31}$\\
$\p_{42}$ &$ -\w \p_{32} $& $\w^2 \p_{12}$ &$ -\w \p_{22} $&$ \w^2 \p_{42} $
&$ \p_{12}$&$ \w^2 \p_{22}$ & $ -\w^2 \p_{32}$\\
\hline
\end{tabular}
\end{table}

\begin{table} \label{MoreTransforms}
\caption{Effect of 2-fold glide rotations on low-energy
eigenstates. }
\begin{tabular}{|c|c|c|c|c|c|c|}
\hline
\hline
 & $\tilde{C}_{12}$ & $\tilde{C}_{13}$ & $\tilde{C}_{14}$ & $\tilde{C}_{23}$ & $\tilde{C}_{24}$ & $\tilde{C}_{34}$ \\
\hline
$\p_{11}$ & $\w \p_{22}$ & $\w^2 \p_{32}$ & $-\w \p_{42}$
& $-\w^2 \p_{12} $&  $- \p_{12}$ & $ \w \p_{12}$ \\
$\p_{12}$ & $\w^2 \p_{21}$ &$ \w \p_{31}$ & $-\w^2 \p_{41}$
&$ -\w \p_{11}$ & $ -\p_{11} $&$ \w^2 \p_{11}$\\
$\p_{21}$ & $\p_{12} $& $\w \p_{22} $&  $\p_{22} $&
$\w \p_{32}  $ & $ -\p_{42}$&$ -\w^2 \p_{22} $\\
$\p_{22}$ & $\p_{11} $& $\w^2 \p_{21}$ & $\p_{21} $&
$\w^2 \p_{31} $& $ - \p_{41}$&$  -\w \p_{22}$ \\
$\p_{31}$ & $-\w \p_{32}$ & $\p_{12} $& $\w \p_{32}$ &
$-\p_{22} $ &$ \p_{32} $&$ -\w^2 \p_{42}$\\
$\p_{32}$ & $-\w^2\p_{31}$ & $\p_{11} $&$ \w^2 \p_{31}$ &
$- \p_{21}$ & $ \p_{31} $&$ -\w \p_{41}$\\
$\p_{41}$ & $-\w^2 \p_{42}$ & $\w \p_{42}$ & $- \w^2 \p_{12}$ &
$  \w \p_{42}$ &$ \p_{22}$&$ \w \p_{32}$ \\
$\p_{42}$ &$ -\w \p_{41} $& $\w^2 \p_{41}$ &$ -\w \p_{11} $&
$ \w^2 \p_{41} $ &$ \p_{21}$&$ \w^2 \p_{31}$\\
\hline
\end{tabular}
\end{table}

\bibliography{PyroMFPaper9}

\end{document}